\documentclass[%
reprint,
superscriptaddress,
prd,
 amsmath,
 amssymb,
 aps,
 longbibliography,
 nofootinbib,
]{revtex4-2}
\usepackage{lipsum, babel}
\usepackage{graphicx} % For including images
\usepackage{float}
\usepackage{cancel}
\usepackage{dcolumn}% Align table columns on decimal point
\usepackage{bm}% bold math
\usepackage[usenames,dvipsnames]{color}
\usepackage{xcolor}
\usepackage{soul}
\usepackage{subcaption}
\usepackage{enumitem}
\usepackage{xspace}
\usepackage{booktabs}
\usepackage{siunitx} % Add support for units
\usepackage[ISO]{diffcoeff}
\usepackage{ragged2e}
\usepackage{mathrsfs}
\usepackage{hyperref}
\usepackage[capitalise]{cleveref}
\usepackage[normalem]{ulem}
\usepackage{wrapfig}
\usepackage{esvect} %for correct vectors using \vv
\usepackage{titlesec}
\usepackage{titletoc}
\usepackage{chngcntr}
\usepackage{fontawesome5}
\usepackage{mleftright}
\usepackage[nodayofweek]{datetime}
\usepackage{comment}
\usepackage{adjustbox}

\DeclareSIUnit\year{yr} 
\DeclareCaptionJustification{justified}{\justifying}
\definecolor{firebrick}{HTML}{B22222}

\hypersetup{
    colorlinks=true,       % false: boxed links; true: colored links
    linkcolor=firebrick,          % color of internal linkshttps://www.overleaf.com/project/65dd05895a901323fe62bed3
    citecolor=firebrick,        % color of links to bibliography
    filecolor=firebrick,      % color of file links
    urlcolor=firebrick           % color of external links
}

\definecolor{orcid-green}{RGB} {166, 206, 57}
\newcommand{\MYhref}[3][blue]{\href{#2}{\color{#1}{#3}}}%

\titleclass{\mysection}{straight}[\section]
\titleformat{\mysection}[runin]
  {\itshape}{\thesection}{}{}[.---]
\titlespacing{\mysection}{1em}{1em}{0em}

\definecolor{newgreen}{HTML}{007A5E}

\DeclareSIUnit\clight{c}

\begin{document}

\title{\boldmath  
 \textbf{{A Unified Numerical Study of Axion Stars:\\
 From the Nonrelativistic Regime to General Relativity}}}

\author{Parisa Arabameri}
\affiliation{Facultad de Ingenier\'ia, Universidad San Sebasti\'an,
Bellavista 7, Santiago 8420524, Chile}

\author{Paola Arias\,\MYhref[orcid-green]{
https://orcid.org/0000-0001-5551-9182}{\faOrcid}}
\affiliation{{Departamento de F\'isica, Universidad T\'ecnica Federico Santa Mar\'ia,
Casilla 110-V, Avda. Espa\~na 1680, Valpara\'iso, Chile}}

\author{Francisco Colip\'i-Marchant\,\MYhref[orcid-green]{
https://orcid.org/0009-0006-3900-0018}{\faOrcid}}
\affiliation{Facultad de Ingenier\'ia, Universidad San Sebasti\'an,
Bellavista 7, Santiago 8420524, Chile}

\author{Enrico D.~Schiappacasse\,\MYhref[orcid-green]{
https://orcid.org/0000-0002-6136-1358}{\faOrcid}}
\affiliation{Facultad de Ingenier\'ia, Universidad San Sebasti\'an,
Bellavista 7, Santiago 8420524, Chile}

\date{\today}

\begin{abstract} 
Axion-star mass--radius relations are commonly computed using different orders of relativistic approximation, making it important to
determine where these descriptions remain reliable. We perform a unified
numerical comparison of axion-star ground-state configurations in the
Newtonian Schr\"odinger--Poisson description, first- and second-order
relativistic effective field theories, and the full
Einstein--Klein--Gordon system for a real scalar field. Using the same
attractive quartic self-interaction in all four descriptions, we scan
\(
|\tilde{\lambda}|=(M_{\rm Pl}/f_a)^2
\)
and determine the maximum masses and corresponding enclosed-mass radii.
All descriptions recover the common large-$|\tilde{\lambda}|$
dilute-star scaling, while substantial differences appear at weak and
moderate coupling. 
The relativistic EFTs interpolate systematically between the
Newtonian and full-GR results. For part of the maximum-mass sequence where $\max|\phi|/f_a=\mathcal{O}(1)$, we test the temporal-harmonic
and potential truncations explicitly in full GR. The higher-harmonic
expansion shows rapid convergence, while restoring the complete
single-cosine potential changes the maximum mass only at the percent
level and $R_{95}$ at the several-percent level. Together with the
systematic convergence of the relativistic EFT descriptions toward full
GR, these results show that the large weak-coupling departure from the
Schr\"odinger--Poisson prediction reflects the breakdown of the
nonrelativistic structural description. Our results provide a
systematic benchmark for determining when Newtonian, relativistically
corrected, or fully general-relativistic descriptions are required for
axion-star structure.
The numerical implementation used in this work is available in the Axion Star Solvers repository at \url{https://github.com/Parisa-Arabameri/AxionStar}.
\end{abstract}
\maketitle

\tableofcontents

%%%%%%%%%%%%%%%%%%%%%%%%%%%%%%%%%%%%%%%%%%%%%%%%%%%%%%%%%%%%%%%%%%%%%%%

\section{Introduction}

Dark matter is one of the clearest indications that our current description of the matter content of the Universe is incomplete. Its existence is strongly supported by gravitational evidence on galactic, cluster, and cosmological scales, but its microscopic nature remains unknown. In particular, dark matter has not yet been detected through nongravitational interactions, and therefore its mass, spin, self-interactions, and couplings to the Standard Model remain open questions. This motivates a broad class of particle physics models beyond the standard cold dark matter paradigm, including light and ultralight bosonic dark matter candidates \cite{Bertone:2004pz,Feng:2010gw,Marsh:2015xka, Arias:2012az}.

Light bosonic dark matter can behave as a coherent classical field on astrophysical scales when it has a sufficiently high phase-space density, $n\,\lambda_{\rm dB}^{3}\gg 1$. For ultralight bosonic candidates in the galactic halo, the occupation number can easily reach $\gtrsim 10^{20}$, so that the field is highly classical. In this regime, gradient pressure, together with scalar self-interactions when present, can balance self-gravity, yielding localized equilibrium configurations known as solitons \cite{Guth:2014hsa}.

In the present context, by a soliton we mean a long-lived, spatially localized, coherent field configuration whose self-gravity is balanced by gradient pressure and, depending on the model, by scalar self-interactions. These solitons are not hard surface compact objects. Rather, they are
smooth, spatially extended field configurations whose macroscopic mass,
radius, and density profile emerge from the collective wave dynamics of
the underlying bosonic field
\cite{Chavanis:2011Analytical,ChavanisDelfini:2011Numerical,
Schiappacasse:2017ham}.

Localized bosonic configurations can form dynamically through gravitational cooling \cite{SeidelSuen:1994GravitationalCooling}, appear as central cores in fuzzy dark matter halos \cite{Schive:2014hza}, and nucleate via gravitational Bose–Einstein condensation in virialized axion miniclusters, minihalos, and PBH-induced minihalos \cite{Levkov:2018kau,Eggemeier:2019khm,Hertzberg:2020hsz,Yin:2024xov}. Analogous localized structures also exist for higher-spin bosons \cite{Jain:2021pnk,Amin:2022pzv}
\footnote{Recent work has explored solitons composed of vector
(spin-1) and tensor (spin-2) dark matter. Vector solitons can exhibit
distinct polarization-dependent structures and can generate
electromagnetic signals through parametric resonance, dipole
interactions, kinetic mixing, and merger events
\cite{Amin:2023imi,Schiappacasse:2026vector,Amaral:2025fcd}.
Spin-2 solitons can likewise source electromagnetic radiation through
dimension-six interactions
\cite{Schiappacasse:2025mao}.}. 

In this work we focus on spin-0 axion dark matter \cite{Arias:2012az}. The axion is a real pseudoscalar field with an approximately periodic self-interaction potential. In the nonrelativistic regime, its slowly varying envelope obeys Schrödinger–Poisson type equations, whose localized solutions are known as axion solitons or axion stars. For attractive self-interactions, the equilibrium sequence has two branches meeting at a maximum-mass turning point: a large-radius branch stable under small radial perturbations, on which gravity dominates the self-interaction, and a small-radius unstable branch on which the self-interaction dominates. Above the turning point, no stable dilute equilibrium exists within the nonrelativistic quartic description, and the configuration may undergo collapse \cite{Schiappacasse:2017ham}. Throughout, ``stable'' and ``unstable'' refer to this nonrelativistic perturbative classification unless a relativistic dynamical analysis is explicitly stated. Because these solitons concentrate dark matter into compact, high-density, coherently oscillating objects, they are relevant for several observational channels.
As an illustrative spin-0
benchmark, consider a dilute axion star at the maximum-mass point
of its stable branch. 
Within the variational treatment of
Refs.~\cite{Schiappacasse:2017ham,Fujikura:2021omw}, the maximum
mass and corresponding minimum characteristic scale are
\begin{align}
M_{\max}
&\simeq
1.5\times10^{-11}M_\odot
\left(\frac{10^{-5}\,{\rm eV}}{m}\right)
\left(\frac{f_a}{10^{12}\,{\rm GeV}}\right)
\left(\frac{0.3}{\gamma}\right)^{1/2},
\\
\label{Rmin-eq}
R_{\min}
&\simeq
1.8\times10^{4}\,{\rm m}
\left(\frac{10^{-5}\,{\rm eV}}{m}\right)
\left(\frac{10^{12}\,{\rm GeV}}{f_a}\right)
\left(\frac{\gamma}{0.3}\right)^{1/2}.
\end{align}
For the representative values
$m=10^{-5}\,{\rm eV}$,
$f_a=6\times10^{11}\,{\rm GeV}$, and
$\gamma=0.3$, where $m$ and $f_a$ are the axion mass and decay constant, respectively, these relations give
\begin{equation}
M_{\max}\simeq9\times10^{-12}M_\odot,
\qquad
R_{\min}\simeq30\,{\rm km}.
\end{equation}
Here \(\gamma\) parametrizes the strength of the quartic axion
self-interaction according to
\begin{equation}
    \lambda_a=-\gamma\frac{m^2}{f_a^2}.
\end{equation}
The expansion of a single cosine axion potential corresponds to
\(\gamma=1\), whereas the zero-temperature QCD axion potential obtained
from chiral perturbation theory gives
\(\gamma=0.346(22)\), which is commonly approximated as
\(\gamma\simeq0.3\)
\cite{diCortona:2015ldu}. 
 In our numerical analysis, we consider $\gamma=1$.

In the
standard post-inflationary QCD axion scenario, a rough estimate based
on the axion population initially contained within a correlated
QCD horizon volume suggests that the fraction of dark matter assembled
into stable spherical axion clumps may be of order
\cite{Guth:2014hsa,Fujikura:2021omw}
\begin{equation}
    \frac{\Omega_{\rm clump}}{\Omega_{\rm DM}}
    \sim 0.1,
\end{equation}
or smaller. This estimate is strongly dependent on the cosmological
history, the formation mechanism, and the angular momentum distribution
of the clumps, and should not be interpreted as a universal prediction.
Axion solitons may nevertheless constitute a subdominant but
phenomenologically important component of the dark matter distribution.

The high internal field amplitude and coherent oscillation of axion stars enhance non-gravitational axion couplings and can drive stimulated photon production through parametric resonance, while their extended mass distribution acts as a gravitational lens, motivating searches through microlensing \cite{Fujikura:2021omw}, axion–photon resonance \cite{Hertzberg:2018zte}, and merger-triggered emission \cite{Hertzberg:2020dbk}.

These signatures depend sensitively on the soliton mass, radius, density
profile, and the critical maximum mass separating the stable and
unstable dilute branches.
Therefore, understanding how relativistic corrections modify the axion star mass--radius relation is necessary for connecting microscopic axion parameters to astrophysical searches. 
The need for a relativistic treatment becomes especially clear when
axion stars approach the maximum-mass turning point or participate in
strongly dynamical processes. Collapse, migration in the strong field
regime, black hole formation, and encounters with compact objects cannot
be described reliably by the purely Newtonian
Schr\"odinger--Poisson approximation. Fully nonlinear
general relativistic simulations have shown that axion-star evolution can
lead to long-lived oscillating configurations, dispersal through
gravitational cooling, or black hole formation, depending on the initial
mass and axion parameters \cite{Helfer:2016ljl}. Axion-star collisions
with black holes and neutron stars have likewise been studied in full
\(3+1\) numerical relativity, where the dynamical spacetime is essential
for consistently following the scalar, gravitational, and matter
 components of the system \cite{Clough:2018exo}.

In addition, compact axion stars coupled to electromagnetism through the axion Chern--Simons interaction can develop an electromagnetic instability whose onset depends not only on the microscopic coupling $g_{a\gamma\gamma}$ but also on the field amplitude and spatial extent of the star \cite{Chung-Jukko:2023wlc}, and constraints from the decay of supercritical axion stars into photons depend on the critical soliton mass and on the abundance of objects exceeding the instability threshold \cite{Escudero:2023vgv}.

These studies show that axion-star phenomenology is not restricted to the dilute nonrelativistic regime. Although an equilibrium mass--radius sequence does not by itself determine the nonlinear outcome of an unstable configuration,  it establishes the maximum mass, radius,
and compactness of the configuration from which the
subsequent evolution begins.

Whether the Newtonian approximation is sufficient depends on the axion
parameters and on the position of the configuration along the
equilibrium sequence. 
Relativistic effects become increasingly important for compact
configurations near the maximum-mass turning point, particularly in the
weak to moderate coupling regime. Reliable structural predictions in
this region therefore require relativistically corrected or fully
general relativistic equilibrium solutions.

In this work, we study axion-star equilibrium sequences from the
Newtonian Schr\"odinger--Poisson regime through first- and second-order
relativistic effective field theory to the full
Einstein--Klein--Gordon system. Our goal is to quantify where the
nonrelativistic approximation remains reliable and how the maximum mass
and enclosed-mass radius change as the attractive self-interaction is
varied. To the best of our knowledge, a systematic numerical comparison
of these four descriptions, using the same attractive quartic potential
and consistent definitions of mass and radius, has not previously been
performed. This comparison identifies the parameter region in which
nonrelativistic structural predictions are adequate and the region in
which relativistically corrected or fully relativistic results should
instead be used.

\section{Relativistic Corrections For Axion Star}

We adopt as our theoretical starting point the nonrelativistic effective field theory (EFT) for a real scalar field $\phi$, as developed in \cite{salehian_beyond_2021}. This EFT is derived by systematically integrating out the relativistic (fast) modes from the full Einstein--Klein--Gordon (EKG) system in a perturbed FLRW background, yielding a set of effective equations that go beyond the standard Schrödinger-Poisson (SP) approximation.

Let us recall the main steps in the derivation of these equations.

The dynamics of a real scalar field minimally coupled to gravity are governed by the relativistic action
\begin{align}
S &= \int d^4x \sqrt{-g} \bigg[ \frac{1}{2} M_{\rm Pl}^2 R - \frac{1}{2} g^{\mu\nu} \partial_\mu \phi \partial_\nu \phi 
- V(\phi) \bigg],
\end{align}

Here \(g\equiv\det(g_{\mu\nu})\), \(g^{\mu\nu}\) is the inverse spacetime
metric, \(R\) is the Ricci scalar, and
\begin{equation}
    M_{\rm Pl}\equiv\frac{1}{\sqrt{8\pi G}}
\end{equation}
is the reduced Planck mass. The function \(V(\phi)\) denotes the full
scalar potential, including both the quadratic mass term and the
self-interaction terms. 
Since the objects considered here are localized bound configurations whose size is much smaller than the Hubble scale, we neglect cosmological expansion in the equilibrium problem and impose asymptotic flatness $g_{\mu\nu}\rightarrow \eta_{\mu\nu}$, 
where \(\eta_{\mu\nu}\) is the Minkowski metric.

For an axion-like field, the shift symmetry of the underlying pseudo-Nambu--Goldstone boson is broken by nonperturbative effects, generating an approximately periodic potential
\cite{Peccei:1977hh,Weinberg:1977ma,Wilczek:1977pj,diCortona:2015ldu,Zhang:2020axionstars}
\begin{align}
    V(\phi)=m^2 f_a^2\left(1-\cos \frac{\phi}{f_a}\right)\approx \frac{1}{2} m^2\phi^2 + \frac{\lambda}{4!}\phi^4,
\end{align}
where the potential expansion is valid in the small field regime.  The quartic coupling is given by
\begin{equation}
    \lambda = -\frac{m^2}{f_a^2}.
    \label{lambda_definition}
\end{equation}
The nonrelativistic EFT expansion is controlled by the hierarchy between the typical physical momentum \(k\) of the bound configuration and the particle mass \(m\). Equivalently, 
\begin{equation} 
v\sim \frac{k}{m}\ll 1, \qquad k\ll m. 
\end{equation} 
This hierarchy separates the fast Compton oscillation scale \(m^{-1}\) from the much slower variation of the nonrelativistic envelope field. Relativistic corrections become important when gradients, gravitational potentials, self-interactions, or binding energies are no longer parametrically small compared with the rest mass scale.
To develop a nonrelativistic theory valid in the limit $v \ll 1$, we follow \cite{salehian_beyond_2021} and decompose the real field $\phi$ into a complex field $\psi$ via
\begin{align}
    \phi(t, \vec{r}) = \frac{1}{\sqrt{2m}} \left( \psi(t, \vec{r}) e^{-imt} + \psi^*(t, \vec{r}) e^{imt} \right),
\end{align}
where $\psi(t,\vec{r})$ is assumed to vary slowly in both time and space compared to the Compton scale set by the mass $m$. However, due to nonlinear interactions and gravitational backreaction, $\psi$ generally includes both slow and fast modes (oscillating near harmonics of $m$). To isolate the physical nonrelativistic dynamics, a time-averaged field $\psi_s(t, \vec{r})$ is defined by applying a smearing procedure, as introduced in the following section, that integrates out rapid oscillations
\begin{align}
    \psi_s(t, \vec{r}) \equiv \langle \psi(t, \vec{r}) \rangle,
\end{align}
where the angular brackets denote a time average over scales $\gg m^{-1}$. The field $\psi_s$ thus represents the slowly varying mode of the complex scalar and becomes the dynamical variable in the effective theory.
\subsection{Power Counting and Slow/Fast Mode Expansion}
\label{subsec:power_counting}
	The full relativistic scalar field oscillates on the Compton time scale \(m^{-1}\). As mentioned before, the nonrelativistic field \(\psi\) is slowly varying only at leading order. Once nonlinear self-interactions and gravity are included, \(\psi\) and the metric variables also contain small fast components. Following \cite{salehian_beyond_2021}, for any dynamical variable \(X\) we define its slow part by a time-smearing operation,
	\begin{equation}
		X_s(t,\mathbf{x})
		\equiv
		\langle X(t,\mathbf{x})\rangle
		=
		\int dt'\,W(t-t')X(t',\mathbf{x}),
		\label{eq:slow_mode_definition}
	\end{equation}
	where the window function averages over the fast time scale \(m^{-1}\), but not over the much longer time scale on which the nonrelativistic configuration evolves. Equivalently, each variable can be decomposed as
	\begin{equation}
		X(t,\mathbf{x})
		=
		\sum_{\nu=-\infty}^{\infty}
		X_\nu(t,\mathbf{x})e^{i\nu mt},
		\qquad
		X_\nu
		=
		\left\langle
		X e^{-i\nu mt}
		\right\rangle .
		\label{eq:mode_decomposition_general}
	\end{equation}
	The slow mode is \(X_s=X_0\), while the modes with \(\nu\neq0\) are called nonzero modes. The coefficients \(X_\nu\) are themselves slowly varying functions; the fast oscillation has been factored out explicitly through \(e^{i\nu mt}\). If \(X\) is real, the modes satisfy \(X_{-\nu}=X_\nu^*\).
	
The nonzero modes are not arbitrary extra degrees of freedom. They are sourced by the slow fields through the nonlinear equations of motion. For example, products such as \(|\psi_s|^2\psi_s\), \(\Phi_s\psi_s\), and the oscillatory factors appearing in the real-field decomposition generate harmonics at \(\nu=\pm2,\pm4,\ldots\). Here \(\Phi_s\) denotes the slowly varying Newtonian-like gravitational potential appearing in the scalar sector of the metric. The EFT procedure consists of solving these \(\nu\neq0\) modes perturbatively and substituting them back into the equations for the slow modes. In this way, the fast modes are integrated out, but their backreaction remains as correction terms in the effective equations. This is the origin of the relativistic corrections to the SP system.
	
	The expansion is controlled by small nonrelativistic parameters. In the notation of Ref.~\cite{salehian_beyond_2021}, one introduces
	\begin{align}
		\epsilon_x
		&\sim
		\left|
		\frac{\nabla^2}{m^2}
		\right|,
		&
		\epsilon_\lambda
		&\sim
		\frac{|\lambda|\phi^2}{m^2},
		&
		\epsilon_g
		&\sim
		|\Phi|
		\sim
		|\Psi|,
		&
		\epsilon_\phi
		&\sim
		\frac{|\phi|}{M_{\rm Pl}}.
		\label{eq:epsilon_parameters_static}
	\end{align}
	For the localized equilibrium problem considered here, cosmological
expansion is neglected. The remaining nonrelativistic expansion
parameters are therefore 
	\begin{equation}
		\epsilon
		=
		\{
		\epsilon_x,\epsilon_\lambda,\epsilon_g,\epsilon_\phi
		\}.
		\label{eq:epsilon_collective}
	\end{equation}
	The Schr\"odinger--Poisson system is the leading nonrelativistic limit.

	This power counting also determines how the nonzero modes are treated. Since the nonzero modes are suppressed relative to the slow modes, they are expanded as
	\begin{equation}
		X_\nu
		=
		X_\nu^{(1)}
		+
		X_\nu^{(2)}
		+\cdots,
		\qquad
		\nu\neq0,
		\label{eq:nonzero_mode_epsilon_expansion}
	\end{equation}
	with
	\begin{equation}
		\frac{X_\nu^{(n)}}{X_s}
		\sim
		O(\epsilon^n).
		\label{eq:nonzero_mode_scaling}
	\end{equation}
	Thus, \(X_\nu^{(1)}\) is the first correction sourced by the slow mode, \(X_\nu^{(2)}\) is the next correction, and so on. The second-order equations therefore contain not only higher powers of the slow fields, but also contributions from the nonzero modes that have been solved for at the appropriate order.

\subsection{First-Order Relativistic Corrections}\label{subsec:first-order}

In the specific case of solitonic (localized) configurations, one considers quasi-stationary, spherically symmetric solutions of the form
\begin{equation}
    \psi_s(t, r) = f(r) e^{i \mu t},
    \label{s_symm_profile}
\end{equation}
where $\mu \ll m$ is the nonrelativistic binding energy, or chemical potential,
shift relative to the rest-mass frequency. Because \(\psi_s\) is a slowly varying field,
the stationary ansatz introduces the additional small parameter
\begin{equation}
    \epsilon_t
    \equiv
    \left|
    \frac{\dot{\psi}_s}{m\psi_s}
    \right|
    =
    \frac{|\mu|}{m}
    \ll1.
\label{eq:epsilon_time}
\end{equation}
For the stationary soliton problem, the complete power counting is thus
\begin{equation}
    \epsilon
    =
    \{
    \epsilon_t,
    \epsilon_x,
    \epsilon_\lambda,
    \epsilon_g,
    \epsilon_\phi
    \}.
\end{equation} 

The gravitational potential is also assumed to be spherically symmetric and static,
\begin{equation}
    \Phi_s(t, r) = \Phi(r),
    \label{s_symm_grav_pot}
\end{equation}
consistent with the symmetry and stationarity of the scalar configuration.

In particular, our analysis begins from Equations (4.2) and (4.3) of \cite{salehian_beyond_2021}, which characterize the structure of quasi-stationary, spherically symmetric solitonic solutions in the nonrelativistic regime, including first-order relativistic corrections. Replacing the ansätze \eqref{s_symm_profile} and \eqref{s_symm_grav_pot} into the EFT equations (specifically Eqs. (3.18) and (3.19) in \cite{salehian_beyond_2021}), and expanding up to first-order relativistic corrections, one arrives at the coupled system (Eqs. (4.2) and (4.3) of \cite{salehian_beyond_2021})
\begin{align}
    & \frac{\nabla^2 f}{2m} - \left(\Phi_s + \frac{\mu}{m}\right)mf - \frac{\lambda f^3}{8 m^2} \nonumber \\
    & \ + \left(3\Phi_s^2 + \frac{4\mu}{m}\Phi_s + \frac{\mu^2}{2m^2}\right)mf \nonumber \\
    & \ + \left(2\Phi_s - \frac{\mu}{m}\right)\frac{\lambda f^3}{8m^2} + \frac{3f^3}{16 M_{\rm Pl}^2} - \frac{\lambda^2 f^5}{768m^5}=0,
    \label{eq_f_dim}
\end{align}
and the equation for $\Phi_s(r)$
\begin{align}
    &\nabla^2 \Phi_s - \left(1-6\Phi_s - 3\frac{\mu}{m} \right)\frac{mf^2}{2M_{\rm Pl}^2} + \frac{\lambda f^4}{16 m^2 M_{\rm Pl}^2} =0.
    \label{eq_phi_dim}
\end{align}
Equation~\eqref{eq_f_dim} generalizes the leading
Schr\"odinger equation by incorporating relativistic corrections to
the kinetic, gravitational, and self-interaction terms. In the general
EFT, the relativistic expansion generates higher order spatial
derivatives, including the familiar correction associated with the
relativistic kinetic energy expansion. For the stationary,
spherically symmetric configurations considered here, these
higher order spatial derivatives are eliminated using the lower order
equations, following ~\cite{salehian_beyond_2021}. Consequently,
Eq.~\eqref{eq_f_dim} remains second order in radial derivatives.
Equation~\eqref{eq_phi_dim} similarly extends the Poisson equation by
including relativistic corrections to the gravitational source.

These two equations define the nonrelativistic EFT including first-order relativistic corrections for axion stars, valid in the regime where gravitational potential and field gradients are small compared to the mass scale, yet allow for nonlinear field configurations beyond the reach of the Schrödinger-Poisson approximation. We adopt them as the theoretical backbone of our analysis.

\begin{figure*}[t]
    \centering
    \includegraphics[width=0.92\textwidth]{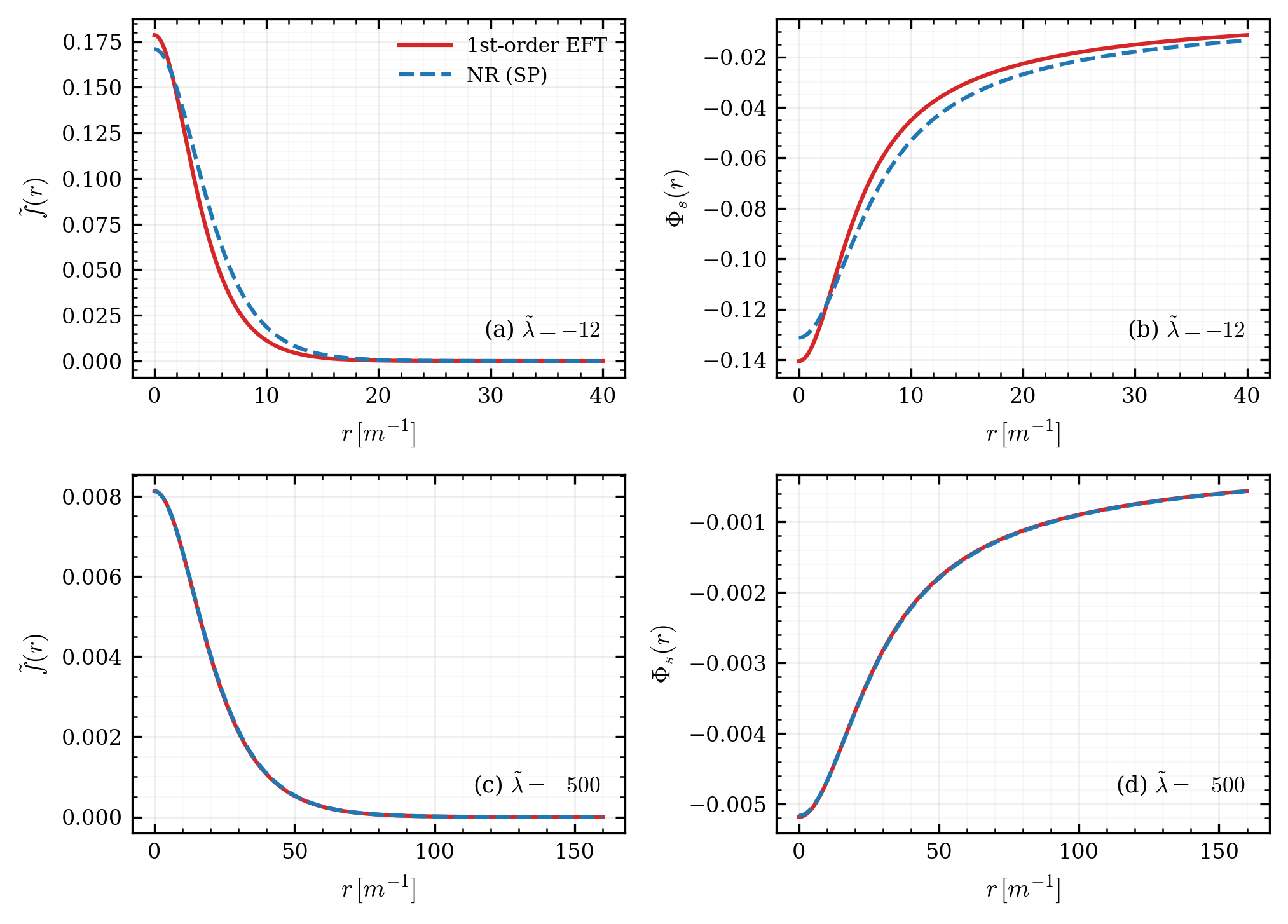}
    \caption{
    Comparison between the Schrödinger--Poisson approximation and the first-order relativistic EFT solution for the axion profile $\tilde f(\tilde r)$ and the gravitational potential $\Phi_s(\tilde r)$. 
    Panels (a) and (b) correspond to $\tilde{\lambda}=-12$, while panels (c) and (d) correspond to $\tilde{\lambda}=-500$. 
    In the $\tilde{\lambda}=-500$ case, the two curves almost overlap, showing that the first-order relativistic correction remains very close to the Schrödinger--Poisson result for the chosen configuration.
}\label{fig:profiles_lambda12_500}
\end{figure*}
%%%%%%%%%%%%
%%%%%%%%%%%%%
\subsection{Equations For Axion Stars} \label{subsec:dimensionless_parameters}
%%%%%%%%%%%%
To solve Equations \eqref{eq_f_dim}-\eqref{eq_phi_dim} numerically, we introduce a suitable set of new variables defined as
\begin{align}
    & r \rightarrow \frac{\tilde{r}}{m}, \quad \mu \rightarrow  \tilde{\mu} m\\ \nonumber
    & f \rightarrow \sqrt{m}M_{\rm Pl} \tilde{f}, \quad \lambda \rightarrow \tilde{\lambda}\frac{m^2}{M_{\rm Pl}^2}. 
\end{align}
This choice leads us to redefine equations  \eqref{eq_f_dim} as
\begin{align}
    & \frac{\tilde{f}''(\tilde{r})}{2} + \frac{\tilde{f}'(\tilde{r})}{\tilde{r}}  - \frac{1}{768} \tilde{\lambda}^2 \tilde{f}(\tilde{r})^5 \nonumber \\
    & - \frac{1}{8} \tilde{\lambda} \tilde{\mu} \tilde{f}(\tilde{r})^3 + \frac{1}{4} \tilde{\lambda} \tilde{f}(\tilde{r})^3 \Phi_s (\tilde{r}) - \frac{1}{8} \tilde{\lambda} \tilde{f}(\tilde{r})^3  \nonumber \\
    &+ \frac{1}{2} \tilde{\mu}^2 \tilde{f}(\tilde{r}) + 4 \tilde{\mu} \tilde{f}(\tilde{r}) \Phi_s (\tilde{r}) - \tilde{\mu} \tilde{f}(\tilde{r}) + 3 \tilde{f}(\tilde{r}) \Phi_s (\tilde{r})^2   \nonumber \\
    & - \tilde{f}(\tilde{r}) \Phi_s (\tilde{r}) + \frac{3 \tilde{f}(\tilde{r})^3}{16} = 0, 
    \label{eq_f_adim}
\end{align}
and \eqref{eq_phi_dim} as 
\begin{align}
    & \Phi_s''(\tilde{r}) + \frac{2\Phi_s'(\tilde{r})}{\tilde{r}} \nonumber \\ 
    & + \tilde{f}(\tilde{r})^2 \left[ 3\Phi_s(\tilde{r}) + \frac{3}{2}\tilde{\mu} - \frac{1}{2} \right] + \frac{1}{16} \tilde{\lambda} \tilde{f}(\tilde{r})^4 = 0, 
    \label{eq_phi_adim}
\end{align}
where $'$ are dimensionless radial derivatives $\partial_{\tilde{r}}$. The leading self-interacting Schr\"odinger--Poisson system is obtained
by retaining only the leading terms according to the EFT power
counting.

Both Equations \eqref{eq_f_adim} and \eqref{eq_phi_adim} are expressed in dimensionless form and are solved numerically for specific $\{\tilde{\lambda}, \tilde{\mu}\}$ values using the algorithm described in Appendix \ref{app:numerics_first_order}.

\subsection{Axion and gravitational potential profiles}

In this section, we compare numerical solutions of Eqs.~\eqref{eq_f_adim} and \eqref{eq_phi_adim} for representative values of the dimensionless self-interaction parameter, $\tilde{\lambda}=-12$ and $\tilde{\lambda}=-500$, considering both the Schrödinger--Poisson (SP) approximation and the effective field theory including first-order relativistic corrections. 
For a fixed value of $\tilde{\mu}$, each solution determines the shape of the axion profile $\tilde f(\tilde r)$ and the gravitational potential $\Phi_s(\tilde r)$. 
These profiles are shown in Fig.~\ref{fig:profiles_lambda12_500}: panels (a) and (b) correspond to $\tilde{\lambda}=-12$, and panels (c) and (d) correspond to $\tilde{\lambda}=-500$. 
The main purpose of these comparisons is to illustrate how relativistic corrections modify the structure of the configurations and how this difference depends on the strength of the self-interaction. 
For \(\tilde{\lambda}=-12\), the SP and first-order EFT profiles are
still close, but small differences are visible near the center. The
first-order solution has a slightly larger central axion amplitude and
a somewhat deeper gravitational potential, corresponding to a more
compact configuration. For \(\tilde{\lambda}=-500\), the two profiles
are nearly indistinguishable, showing the expected recovery of the
nonrelativistic limit. Although these profile differences are small,
they can still lead to noticeable differences in the integrated mass
and in \(R_{95}\).
The mass quoted throughout this work is the ADM mass, namely the gravitational mass measured by an observer at spatial infinity. In the EFT calculations it is obtained from the mass functional associated with the scalar profile and metric potentials, while in the full-GR calculation discussed in the next section, it is extracted from the asymptotic behavior of the metric. Since axion stars are smooth field configurations rather than objects with a sharp surface, they do not possess a hard radius. We therefore characterize their size using the effective radius $R_{95}$, defined as the radius enclosing $95\%$ of the total mass,
\begin{equation}
M(r<R_{95})=0.95\,M_s.
\end{equation}
This definition gives a practical measure of the spatial extent of the configuration and allows a consistent comparison between the Schrödinger--Poisson, relativistic-correction, and full-GR solutions. Details of the mass and radius extraction are given in Appendix~\ref{app:numerics}.

\subsection{Mass--Radius Relation}

To analyze the effect of relativistic corrections, we construct the mass--radius relation from the numerical profile solutions and the corresponding ADM mass defined in the Appendix~\ref{app:numerics_first_order}, equation \eqref{soliton_mass}, for different values of $\tilde{\lambda}$. 
This relation allows us to characterize the family of equilibrium configurations associated with a given pair $\{\tilde{\lambda},\tilde{\mu}\}$.

\begin{figure}[t]
    \centering
    \includegraphics[width=\linewidth]{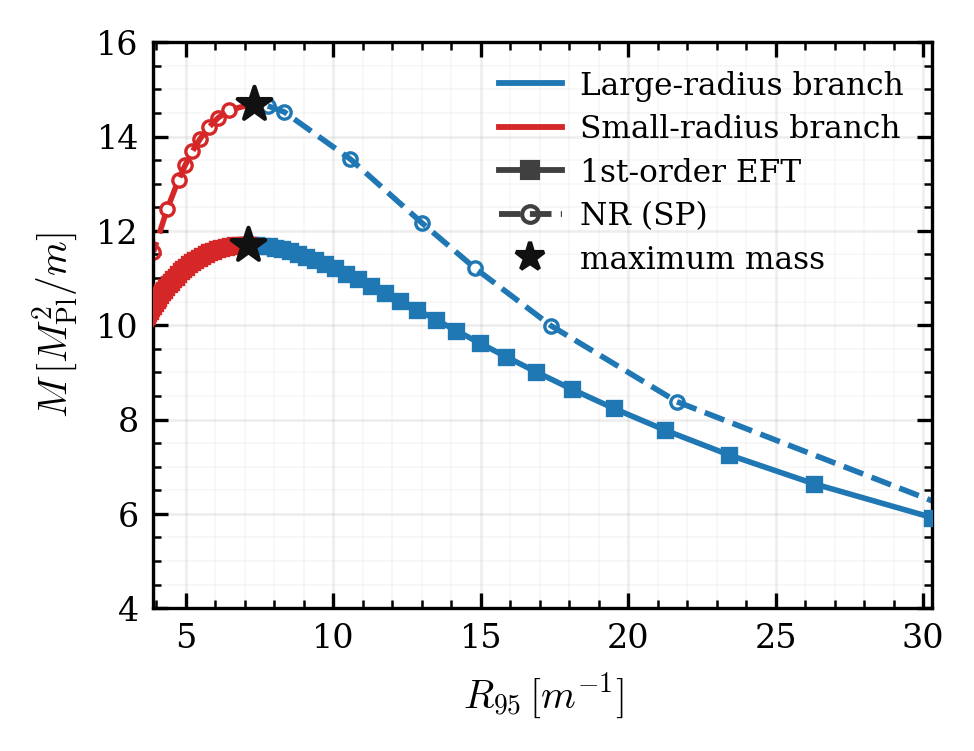}
    \caption{
    Mass--radius relation for \(\tilde{\lambda}=-12\), comparing the Schrödinger--Poisson (SP) and first-order relativistic EFT solutions. Blue and red portions denote the large- and small-radius branches, respectively, while stars mark the maximum-mass turning points. In the SP description, the large-radius branch is stable and the small-radius branch unstable under small radial perturbations.}
\label{fig:mass_radius_branches_lambda12}
\end{figure}

As a representative example, Fig.~\ref{fig:mass_radius_branches_lambda12} compares the mass--radius sequences for \(\tilde{\lambda}=-12\) in the SP approximation and in the first-order relativistic EFT.  
{As discussed in the Introduction, the SP sequence contains two
equilibrium branches separated by a maximum-mass turning point. In the
SP description, the large-radius and small-radius branches correspond,
respectively, to configurations that are stable and unstable under
small radial perturbations \cite{Schiappacasse:2017ham}.} 
{From a phenomenological point of view, we are mainly interested in the
large-radius side of this sequence and in the location of the
maximum-mass turning point. The first-order EFT sequence shown
in Fig.~\ref{fig:mass_radius_branches_lambda12} preserves the same
qualitative two-branch morphology. {For oscillatons, the turning-point criterion has been verified
numerically in the noninteracting limit~\cite{Alcubierre2003}.}

Configurations driven beyond such a critical configuration may
subsequently migrate, disperse, radiate relativistic particles, or
collapse, depending on the scalar potential and on the nonlinear
dynamics
\cite{Chavanis:2017loo,Helfer:2016ljl}.} 

\subsection{Second-Order Relativistic Corrections}
Equations~\eqref{eq_f_adim} and \eqref{eq_phi_adim} correspond to the first relativistic correction to the Schr\"odinger--Poisson system. At this order, the two scalar metric potentials appearing in the relativistic EFT can be identified up to higher-order terms, which is why the system can be written in terms of the single gravitational potential \(\Phi_s\). To extend the calculation to the next order in the nonrelativistic expansion, this identification is no longer sufficient. The \(O(\epsilon^2)\) system keeps the two scalar potentials separately and also contains the leading backreaction of the fast oscillatory modes that were integrated out in the EFT procedure.

For the spherically symmetric soliton, vector and tensor metric perturbations vanish by symmetry, and the relevant scalar part of the metric can be written in isotropic coordinates as
	\begin{equation}
		ds^2
		=
		-e^{2\Phi_s}dt^2
		+
		e^{-2\Psi_s}
		\left(
		dr^2+r^2d\Omega^2
		\right).
		\label{eq:isotropic_metric_eft_second_order}
	\end{equation}
	where \(\Phi_s\) is the lapse potential and \(\Psi_s\) is the spatial-curvature potential. At first order, the difference between the two scalar potentials, \(\Psi_s-\Phi_s\), contributes only beyond the order retained in Eqs.~\eqref{eq_f_adim}--\eqref{eq_phi_adim}, so the system can be written using a single gravitational potential. At the next order, however, \(\Phi_s\) and \(\Psi_s\) must be kept separately. This is why the \(O(\epsilon^2)\) equations will contain both potentials.

It is also convenient to remove the explicit chemical-potential parameter from the stationary equations. Following ~\cite{salehian_beyond_2021}, we redefine
	\begin{align}
		f_{\rm old}(r)
		&=
		\left(
		1-\frac{\mu}{2m}
		\right)
		f(r),
		\\
		\Phi_{s,{\rm old}}(r)
		&=
		\widehat{\Phi}_s(r)
		-
		\frac{\mu}{m}
		-
		\frac{\mu^2}{2m^2}
		-
		\frac{\mu^3}{3m^3}.
		\label{eq:mu_removed_redefinition}
	\end{align}
	After this redefinition, the second-order equations are written in terms of the shifted lapse potential \(\widehat{\Phi}_s\) and the spatial potential \(\Psi_s\). The difference \(\widehat{\Phi}_s-\Psi_s\) is now of the same order as the binding-energy parameter and must be retained.

For a spherically symmetric function \(X(\tilde r)\), we define
    \begin{equation}
		\Delta X
		\equiv
		X''
		+
		\frac{2}{\tilde r}X',
		\label{eq:radial_laplacian_second_order}
	\end{equation}
where primes denote derivatives with respect to the dimensionless radius \(\tilde r\). The unknown functions in the \(O(\epsilon^2)\) system are
\begin{equation}
	\tilde f(\tilde r),
	\quad
	\widehat{\Phi}_s(\tilde r),
	\quad
	\Psi_s(\tilde r),
	\quad
	\Phi^{(1)}_2(\tilde r),
	\quad
	\Psi^{(2)}_2(\tilde r).
\label{eq:second_order_unknowns}
\end{equation}
The last two functions are the nonzero mode metric corrections that survive at this order. The subscript \(2\) indicates that they multiply the second harmonic of the fast oscillation, while the superscripts indicate their order in the \(\epsilon\) expansion. Physically, they represent the leading imprint of the oscillatory metric response of the underlying real scalar field on the time averaged EFT description.

The second-order corrected scalar profile equation is
\begin{align}
	0={}&
	\frac{1}{2}\Delta\tilde f
	-
	\left(
	1-\widehat{\Phi}_s-2\Psi_s
	+2\Psi_s^2
	+2\Psi_s\widehat{\Phi}_s
	+\frac{2}{3}\widehat{\Phi}_s^2
	\right)
	\widehat{\Phi}_s\tilde f\nonumber\\
	&
	+3\Psi^{(2)}_2\tilde f+
	\nonumber
	\left(
	1-2\Psi_s-\frac{4}{3}\widehat{\Phi}_s
	\right)
	\frac{3\tilde f^3}{16}\nonumber\\
	&
	+
	\frac{1}{2}
	\tilde f'
	\left(
	\widehat{\Phi}_s-\Psi_s+\Phi^{(1)}_2
	\right)'\nonumber\\
	&
	-
	\frac{\tilde f(\tilde f')^2}{16}-
	\left(
	1-2\Psi_s+2\Psi_s^2+\Phi^{(1)}_2-\frac{\tilde f^2}{8}
	\right)
	\frac{\tilde\lambda\tilde f^3}{8}\nonumber\\
	&
	-
	\left(
	1-2\Psi_s+\frac{3}{2}\widehat{\Phi}_s
	\right)
	\frac{\tilde\lambda^2\tilde f^5}{768}
	+
	\frac{\tilde\lambda^2\tilde f^3(\tilde f')^2}{1024}
	+
	\frac{\tilde\lambda^3\tilde f^7}{73728}.
	\label{eq_f_second_order}
\end{align}

The equation for the spatial metric potential is
\begin{align}
	0={}&
	\Delta\Psi_s
	-\frac{1}{2}(\Psi_s')^2-\nonumber\\
	&
	(
	1-\widehat{\Phi}_s-2\Psi_s
	+2\Psi_s^2
	+2\widehat{\Phi}_s\Psi_s
	+\widehat{\Phi}_s^2
	+\Phi^{(1)}_2
	-\frac{3\tilde f^2}{32}
	)
	\frac{\tilde f^2}{2}
	\nonumber\\
	&-
	\frac{(\tilde f')^2}{4}
	-
	(1-2\Psi_s)
	\frac{\tilde\lambda\tilde f^4}{32}
	-
	\frac{13\tilde\lambda^2\tilde f^6}{18432}.
	\label{eq_psi_second_order}
\end{align}
The equation for the shifted lapse potential is
\begin{align}
	0={}&
	\Delta\widehat{\Phi}_s
	-\nonumber\\
	&
	(
	1-4\widehat{\Phi}_s-2\Psi_s
	+2\Psi_s^2
	+8\widehat{\Phi}_s\Psi_s
	+4\widehat{\Phi}_s^2
	+\Phi^{(1)}_2\nonumber\\
	&
	+\frac{3\tilde f^2}{16}
	)
	\frac{\tilde f^2}{2}
	+
	\widehat{\Phi}_s'
	(
	\widehat{\Phi}_s-\Psi_s
	)'\nonumber\\
	&
	+
	(1-2\Psi_s)
	\frac{\tilde\lambda\tilde f^4}{16}
	-
	\frac{\tilde\lambda^2\tilde f^6}{18432}.
	\label{eq_phi_second_order}
\end{align}
The nonzero-mode corrections are fixed by the constraints
\begin{equation}
	\left(\Psi^{(2)}_2\right)'
	=
	\frac{\tilde f^2}{16}\widehat{\Phi}_s',
	\label{eq_psi2_constraint}
\end{equation}
and
\begin{equation}
	\Delta\Phi^{(1)}_2
	=
	12\Psi^{(2)}_2
	-
	\frac{\tilde f^2}{2}\widehat{\Phi}_s
	-
	\frac{\tilde\lambda\tilde f^4}{64}.
	\label{eq_phi2_constraint}
\end{equation}
Equations~\eqref{eq_f_second_order}--\eqref{eq_phi2_constraint} are the second-order extension of Eqs.~\eqref{eq_f_adim}--\eqref{eq_phi_adim}. The first-order system corrects the SP balance using a single Newtonian-like potential, while the second-order system resolves the distinction between the two scalar metric potentials and includes the leading effect of the nonzero harmonics induced by the real scalar nature of the field.

At this order, the mass functional must also be corrected. In the same dimensionless convention, the soliton mass is
\begin{align}
	M_s
	=
	4\pi\int_0^\infty d\tilde r\,\tilde r^2
	\Bigg[
	&
	\tilde f^2
	(
	1-\widehat{\Phi}_s-\frac{5}{2}\Psi_s
	+\frac{25}{8}\Psi_s^2
	+\frac{5}{2}\widehat{\Phi}_s\Psi_s\nonumber\\
	&
	+\widehat{\Phi}_s^2
	+\Phi^{(1)}_2
	-\frac{3\tilde f^2}{32}
	)\nonumber\\
	&
	+
	\frac{(\tilde f')^2}{2}
	(
	1-\frac{1}{2}\Psi_s
	)\nonumber\\
	&
	+
	\frac{\tilde\lambda\tilde f^4}{16}
	(
	1-\frac{5}{2}\Psi_s
	)
	+
	\frac{13\tilde\lambda^2\tilde f^6}{9216}
	\Bigg].
	\label{eq_mass_second_order}
\end{align}
The radius \(R_{95}\) is then defined using the same prescription as in the first-order calculation, namely by the condition
\begin{equation}
	M(\tilde r<R_{95})=0.95M_s.
	\label{eq_R95_second_order}
\end{equation}

\section{Relation to the Full General Relativistic Real Field Problem
\label{subsec:relation_full_GR}}

The EFT systems discussed above are obtained by integrating out the fast oscillatory modes of the real scalar field. This distinction is important because the axion is a spin-0 pseudoscalar particle represented by a real scalar field. The complex field used in the nonrelativistic EFT is not an additional fundamental degree of freedom; it is an envelope field that captures the slowly varying amplitude and phase of the real scalar oscillation after the fast Compton-scale dependence has been factored out. The full-GR oscillaton calculation keeps the real field explicitly and therefore provides the appropriate relativistic benchmark for the EFT description \cite{Zhang:2020axionstars,Alcubierre2003,salehian_beyond_2021}. To clarify the connection with the fully relativistic calculation, it is useful to compare our notation with the standard oscillaton formalism.

In full GR, the real scalar field should be evolved directly, without first averaging over the fast oscillations. We write the scalar potential as
\begin{equation}
	V(\phi_{\rm GR})
	=
	\frac{1}{2}m^2\phi_{\rm GR}^2
	+
	\frac{\lambda}{4!}\phi_{\rm GR}^4,
	\label{eq:full_GR_quartic_potential_dim}
\end{equation}
or, after using the dimensionless variables employed in the numerical system,
\begin{equation}
	U(\phi_{\rm GR})
	=
	\frac{1}{2}\phi_{\rm GR}^2
	+
	\frac{\tilde\lambda}{4!}\phi_{\rm GR}^4.
	\label{eq:full_GR_quartic_potential_adim}
\end{equation}

For axion-like particles, {\(\tilde \lambda<0\)}, corresponding to the attractive quartic term obtained by expanding the cosine potential. The quadratic case studied in the standard oscillaton literature \cite{Alcubierre2003} is recovered by taking \(\tilde\lambda=0\).

It is important to emphasize that
Eq.~\eqref{eq:full_GR_quartic_potential_adim} represents the quartic
truncation of the periodic axion potential, rather than the complete
axion potential. The full-GR calculation presented here is therefore
fully relativistic with respect to the gravitational dynamics, while
the scalar self-interaction is treated within the quartic
approximation. Consequently, our results should be interpreted as
applying to a real scalar field with an attractive quartic
self-interaction. {For the baseline comparison among the SP, EFT, and full-GR
descriptions, we therefore use this common quartic interaction.
The accuracy of this truncation for an axion interpretation is
assessed explicitly in Sec.~\ref{sec:quartic_validity}, where we
monitor the real-field excursion and repeat the full-GR calculation
using the complete cosine potential.} 

Following the usual oscillaton construction, we use polar-areal coordinates,
\begin{equation}
	ds^2
	=
	-\alpha^2(t,x)dt^2
	+
	a^2(t,x)dx^2
	+
	x^2d\Omega^2,
	\label{eq:polar_areal_metric_quartic}
\end{equation}
where \(\alpha(t,x)\) is the lapse and \(a(t,x)\) is the radial metric function. The coordinate \(x\) is the areal radius, so that a sphere at fixed \(x\) has area \(4\pi x^2\). This is different from the isotropic radial coordinate used in the EFT metric. Therefore, the metric potentials \(\Phi_s\) and \(\Psi_s\) in the EFT should not be identified directly with \(\alpha\) and \(a\) without a coordinate transformation.

We introduce the variables
\begin{equation}
	\Pi
	=
	\frac{a}{\alpha}\partial_t\phi_{\rm GR},
	\qquad
	\Xi
	=
	\partial_x\phi_{\rm GR}.
	\label{eq:full_GR_first_order_variables_quartic}
\end{equation}
Here \(\Xi\) denotes the radial derivative of the scalar field. We avoid the symbol \(\Psi\) for this derivative because \(\Psi_s\) is already used for the EFT spatial metric potential in the second-order system.

In terms of these variables, the scalar-field energy density and radial pressure are
\begin{align}
	\rho_{\phi}
	&=
	\frac{1}{2a^2}
	\left(
	\Pi^2+\Xi^2
	\right)
	+
	U(\phi_{\rm GR}),
	\label{eq:rho_quartic_full_GR}
	\\
	p_{r,\phi}
	&=
	\frac{1}{2a^2}
	\left(
	\Pi^2+\Xi^2
	\right)
	-
	U(\phi_{\rm GR}).
	\label{eq:pr_quartic_full_GR}
\end{align}
The quartic self-interaction enters the Einstein equations through \(U(\phi_{\rm GR})\), and it enters the Klein--Gordon equation through
\begin{equation}
	\frac{dU}{d\phi_{\rm GR}}
	=
	\phi_{\rm GR}
	+
	\frac{\tilde\lambda}{6}\phi_{\rm GR}^3.
	\label{eq:dU_dphi_quartic}
\end{equation}

With the same polar-areal structure used in the quadratic oscillaton calculation, the generalized quartic EKG system can be written as
\begin{align}
	\frac{a_{,x}}{a}
	&=
	\frac{1-a^2}{2x}
	+
	\frac{x}{4}
	\left[
	\Xi^2+\Pi^2
	+
	2a^2U(\phi_{\rm GR})
	\right],
	\label{eq:full_GR_hamiltonian_quartic}
	\\
	\frac{\alpha_{,x}}{\alpha}
	&=
	\frac{a_{,x}}{a}
	+
	\frac{a^2-1}{x}
	-
	2xa^2U(\phi_{\rm GR}),
	\label{eq:full_GR_slicing_quartic}
	\\
	\partial_t\phi_{\rm GR}
	&=
	\frac{\alpha}{a}\Pi,
	\label{eq:full_GR_phi_evolution_quartic}
	\\
	\partial_t\Pi
	&=
	\frac{1}{x^2}
	\left(
	\frac{x^2\alpha\Xi}{a}
	\right)_{,x}
	-
	a\alpha
	\left(
	\phi_{\rm GR}
	+
	\frac{\tilde\lambda}{6}\phi_{\rm GR}^3
	\right),
	\label{eq:full_GR_pi_evolution_quartic}
	\\
	\partial_t\Xi
	&=
	\left(
	\frac{\alpha\Pi}{a}
	\right)_{,x}.
	\label{eq:full_GR_xi_evolution_quartic}
\end{align}
Equations~\eqref{eq:full_GR_hamiltonian_quartic}--\eqref{eq:full_GR_xi_evolution_quartic} reduce to the usual \(\Phi^2\)-oscillaton equations when \(\tilde\lambda=0\). The only changes introduced by the quartic axion self-interaction are the replacement of the quadratic potential by \(U(\phi_{\rm GR})\) in the metric constraints and the replacement of the mass term in the Klein--Gordon equation by \(dU/d\phi_{\rm GR}\).

The equilibrium construction follows the same logic as in the quadratic oscillaton case. One defines
\begin{equation}
	A(t,x)=a^2(t,x),
	\qquad
	C(t,x)=\left(\frac{a(t,x)}{\alpha(t,x)}\right)^2,
	\label{eq:AC_full_GR_quartic}
\end{equation}
and expands the scalar and metric variables in Fourier modes,
\begin{align}
	\phi_{\rm GR}(t,x)
	&=
	\sum_{j=1}^{j_{\rm max}}
	\phi_j(x)\cos(j\omega t),
	\label{eq:full_GR_scalar_fourier_quartic}
	\\
	A(t,x)
	&=
	\sum_{j=0}^{j_{\rm max}}
	A_j(x)\cos(j\omega t),
	\label{eq:full_GR_A_fourier_quartic}
	\\
	C(t,x)
	&=
	\sum_{j=0}^{j_{\rm max}}
	C_j(x)\cos(j\omega t).
	\label{eq:full_GR_C_fourier_quartic}
\end{align}
The frequency \(\omega\) is determined as part of the eigenvalue problem by requiring regularity at the origin and asymptotic flatness at large radius.

The harmonic structure remains the same as in the quadratic real-field problem because the potential in Eq.~\eqref{eq:full_GR_quartic_potential_adim} is even under \(\phi_{\rm GR}\rightarrow-\phi_{\rm GR}\). Therefore, the scalar field contains odd harmonics,
\begin{equation}
	\phi_{\rm GR}(t,x)
	=
	\phi_1(x)\cos(\omega t)
	+
	\phi_3(x)\cos(3\omega t)
	+\cdots
	\label{eq:full_GR_odd_harmonics_quartic}
\end{equation}
while the metric functions contain even harmonics,
\begin{equation}
	A(t,x)
	=
	A_0(x)
	+
	A_2(x)\cos(2\omega t)
	+
	A_4(x)\cos(4\omega t)
	+\cdots
	\label{eq:full_GR_even_harmonics_quartic}
\end{equation}
and similarly for \(C(t,x)\). The quartic term changes the radial equations for the Fourier coefficients and shifts the equilibrium sequence, but it does not change the parity structure. Physically, this is because the stress-energy tensor is invariant under \(\phi_{\rm GR}\rightarrow-\phi_{\rm GR}\), so the geometry repeats after half a scalar-field period.

For the production full-GR sequences presented in this work, we retain
the first two odd scalar harmonics, together with the metric harmonics through \(4\omega\). This choice is
motivated in part by the multi-harmonic analysis of
~\cite{Visinelli:2017ooc}, which shows that in the dense regime where the characteristic central amplitude of the dimensionless axion field becomes order unity, the single-frequency approximation can fail and higher odd temporal harmonics must be included.
{We therefore do not use a single-harmonic approximation for the
production full-GR calculation. At the same time, the results of
~\cite{Visinelli:2017ooc} do not imply that a fixed two-harmonic
truncation is sufficient in general. We test this explicitly by
enlarging the scalar basis through \(\phi_{11}\), with the metric basis
extended consistently through \(12\omega\). The resulting convergence
of the structural observables and Fourier coefficients is presented in
Appendix~\ref{app:harmonic_convergence}.} {Although this convergence demonstrates that the selected Fourier basis accurately models the periodic core and its structural properties, it remains insufficient for assessing the magnitude of any outgoing radiation or the actual lifespan of the real-field configuration   \cite{Hertzberg_2010_radiative}. Ascertaining these dynamic characteristics would dictate the use of a fully time-dependent analysis subject to radiative boundary conditions.}

\section{Maximum-Mass Scaling and Accuracy}
\begin{figure*}[t]
    \centering
    \begin{subfigure}{0.48\textwidth}
        \centering
        \includegraphics[width=\linewidth]{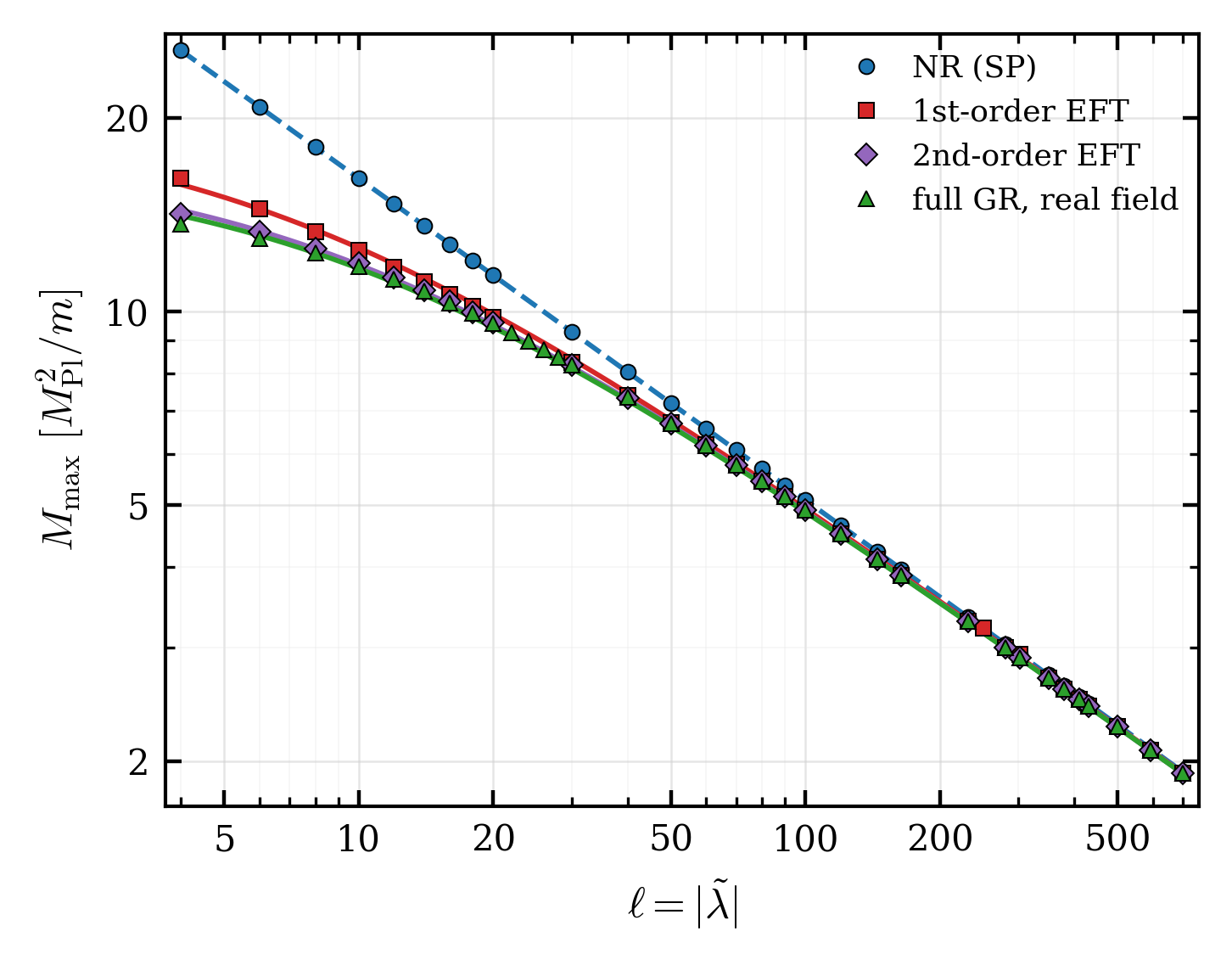}
        \caption{}
        \label{fig:mass_lambda_scaling}
    \end{subfigure}
    \hfill
    \begin{subfigure}{0.48\textwidth}
        \centering
        \includegraphics[width=\linewidth]{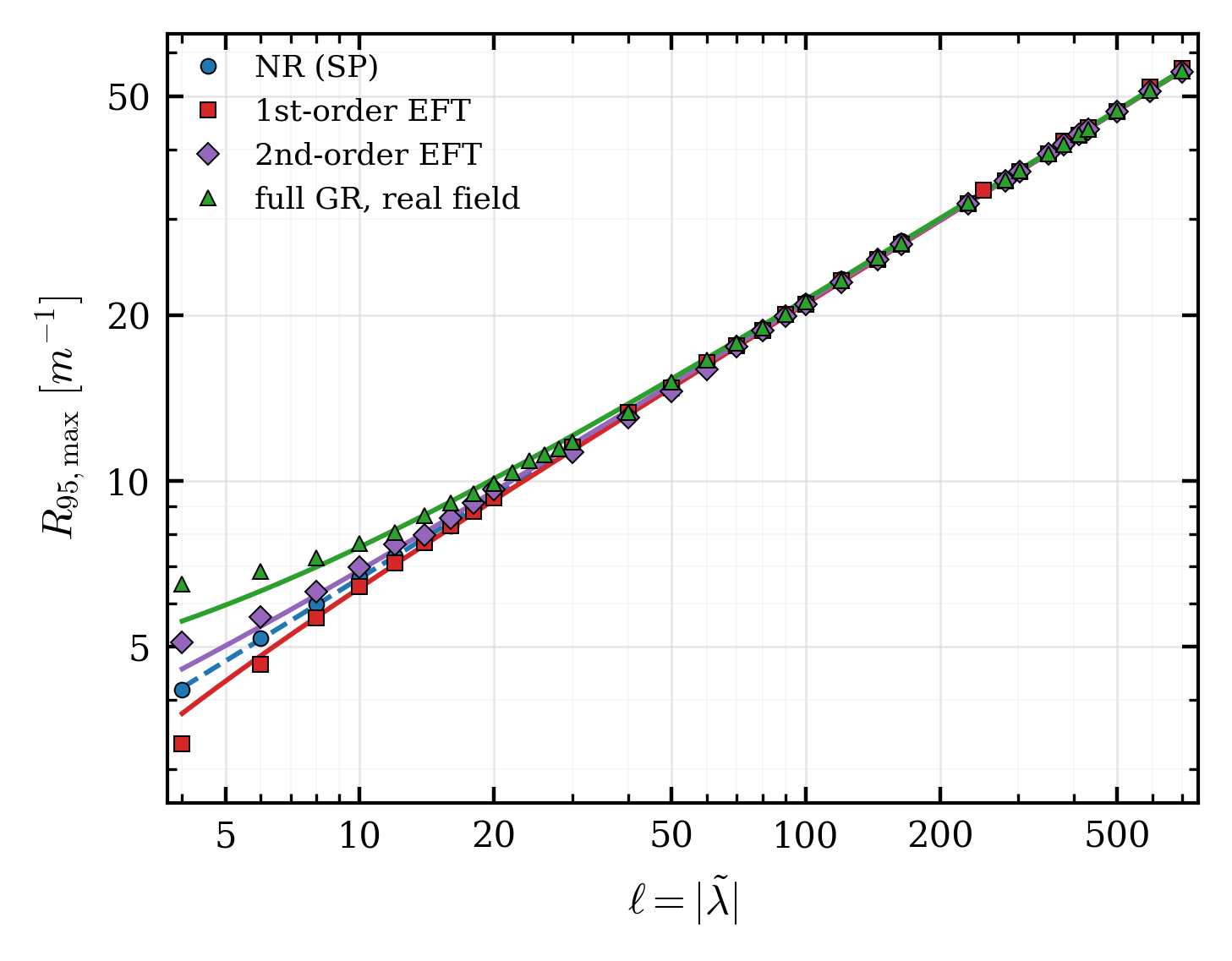}
        \caption{}
        \label{fig:radius_lambda_scaling}
    \end{subfigure}
    \caption{
    Scaling of the maximum mass configurations with the attractive self-interaction parameter \(|\tilde{\lambda}|\). 
    Panel (a) shows the maximum ADM mass \(M_{\max}\) as a function of \(|\tilde{\lambda}|\), together with the fitted relations. 
    Panel (b) shows the corresponding effective radius \(R_{95,\max}\), fitted with the pivoted square root form of Eq.~\eqref{eq:radius-fit-pivot}. 
    The points are numerical maximum-mass configurations and the curves are
    the constrained fits calibrated over \(4\leq|\tilde{\lambda}|\leq700\). The mass
    and radius constraints enforce the common SP large-\(|\tilde{\lambda}|\) asymptotes.
    }
    \label{fig:max_scaling_lambda}
\end{figure*}

To make the comparison between the different approximations more
quantitative, we fit the maximum mass and the corresponding effective
radius as functions of \(
    |\tilde{\lambda}|
\).
All fitting relations quoted below are calibrated over the numerical
interval
\[
    4\leq|\tilde{\lambda}|\leq700,
\]
and should therefore be regarded as interpolation formulas over this
range.

For the Newtonian Schr\"odinger--Poisson sequence, we impose the
theoretical dilute star scaling
\begin{equation}
    M_{\max}^{\rm NR}(|\tilde{\lambda}|)
    =
    A_{\rm NR}|\tilde{\lambda}|^{-1/2},
    \qquad
    A_{\rm NR}=50.893.
\end{equation}
For each relativistic sequence
\(X=\{\text{1st-order EFT},\text{2nd-order EFT},\text{full GR}\}\),
we use
\begin{equation}
    M_{\max}^{X}(|\tilde{\lambda}|)
    =
    \frac{M_0^X}
    {\sqrt{1+C_X|\tilde{\lambda}|}},
    \qquad
    C_X=
    \left(\frac{M_0^X}{A_{\rm NR}}\right)^2 .
    \label{dilute-axion-star-behavior}
\end{equation}
Thus, only \(M_0^X\) is independently fitted for each relativistic
sequence. By construction, all descriptions recover the same
large-\(|\tilde{\lambda}|\) behavior,
\begin{equation}
    M_{\max}^{X}\simeq A_{\rm NR}|\tilde{\lambda}|^{-1/2}.\label{dilute-axion-star-behavior_2}
\end{equation}

A closely related interpolation was discussed in ~\cite{Chavanis:2022fvh} for relativistic self-gravitating Bose–Einstein condensates with attractive \(|\phi|^4\) self-interaction. This reference uses the complex scalar field boson star limit to fix the weak coupling normalization. In the present work, our full-GR sequence corresponds to real scalar field oscillaton solutions. Here, the fitting form is used only over
the numerical interval stated above. In particular, although the
functional form formally approaches \(M_0^X\) as \(|\tilde{\lambda}|\to0\),
the fitted parameter \(M_0^X\) should not be interpreted as an
independent determination of the physical zero-coupling maximum mass. As an independent check, we directly solve the full-GR real scalar
system at \(|\tilde{\lambda}|=0\) using the same Fourier truncation employed for
the production sequences. We obtain
\begin{equation}
    M_{\max}^{\rm GR}(0)\simeq15.138,
\end{equation}
or, in terms of the non-reduced Planck mass, $M_P$,
\begin{equation}
    \frac{M_{\max}^{\rm GR}(0)}{8\pi}
    \simeq0.602\,\frac{M_P^2}{m}.
\end{equation}

This is within approximately one percent of the standard quadratic
oscillaton value
\(M_{\max}\simeq0.606\!-\!0.607\,M_P^2/m\)
\cite{Alcubierre2003,Chavanis:2022fvh}. As a convergence check, extending the scalar harmonics through
\(\phi_5\) and the metric harmonics through \(6\omega\) gives
\(M_{\max}^{\rm GR}(0)/(8\pi)\simeq0.606\,M_P^2/m\), a change of
only \(0.59\%\).

For the radius, the NR sequence is described by
\begin{equation}
    R_{95,\max}^{\rm NR}(|\tilde{\lambda}|)
    =
    B_{\rm NR}|\tilde{\lambda}|^{1/2},
    \quad
    B_{\rm NR}=2.106 .
    \label{eq:radius-fit-pivot_NR}
\end{equation}
For the relativistic sequences we retain the pivoted form
\begin{align}
    \label{eq:radius-fit-pivot}
    R_{95,\max}^{X}(|\tilde{\lambda}|)
    &=
    R_\star^X
    \sqrt{
        1+
        D_X(|\tilde{\lambda}|-|\tilde{\lambda}|_\star)
    },
    \\
    \nonumber
    D_X=&
    \left(\frac{B_{\rm NR}}{R_\star^X}\right)^2,
    \qquad
    |\tilde{\lambda}|_\star=100 .
\end{align}

Only \(R_\star^X\) is therefore independently fitted. This constraint
ensures that all radius sequences recover the common asymptotic behavior
\(R_{95,\max}^{X}\simeq B_{\rm NR}|\tilde{\lambda}|^{1/2}\).
The numerical coefficients are listed in
Appendix~\ref{app:fit}.

The numerical maximum-mass points and the corresponding fits are shown in Fig.~\ref{fig:max_scaling_lambda}. 
Panel (a) shows the maximum ADM mass as a function of the attractive coupling strength, while panel (b) shows the effective radius \(R_{95,\max}\) of the same maximum mass configurations\footnote{The convergence toward the full-GR mass--radius relation is not necessarily monotonic order by order. At moderate coupling, first-order corrections can over-contract the maximum-mass configuration, reducing both $M_{\max}$ and $R_{95,\max}$.  Additional second-order terms, including the distinction between the two scalar metric
potentials, the leading backreaction of the nonzero metric harmonics,
nonlinear gradient terms, and higher self-interaction contributions.
These terms rebalance the solution and shift the radius back toward the
full-GR result. This behavior is discussed in
Sec.~\ref{sec:quartic_validity} in relation to the breakdown of the
nonrelativistic power counting at low $|\tilde{\lambda}|$.}.

\begin{figure}[t]
    \centering
    \includegraphics[width=0.48\textwidth]{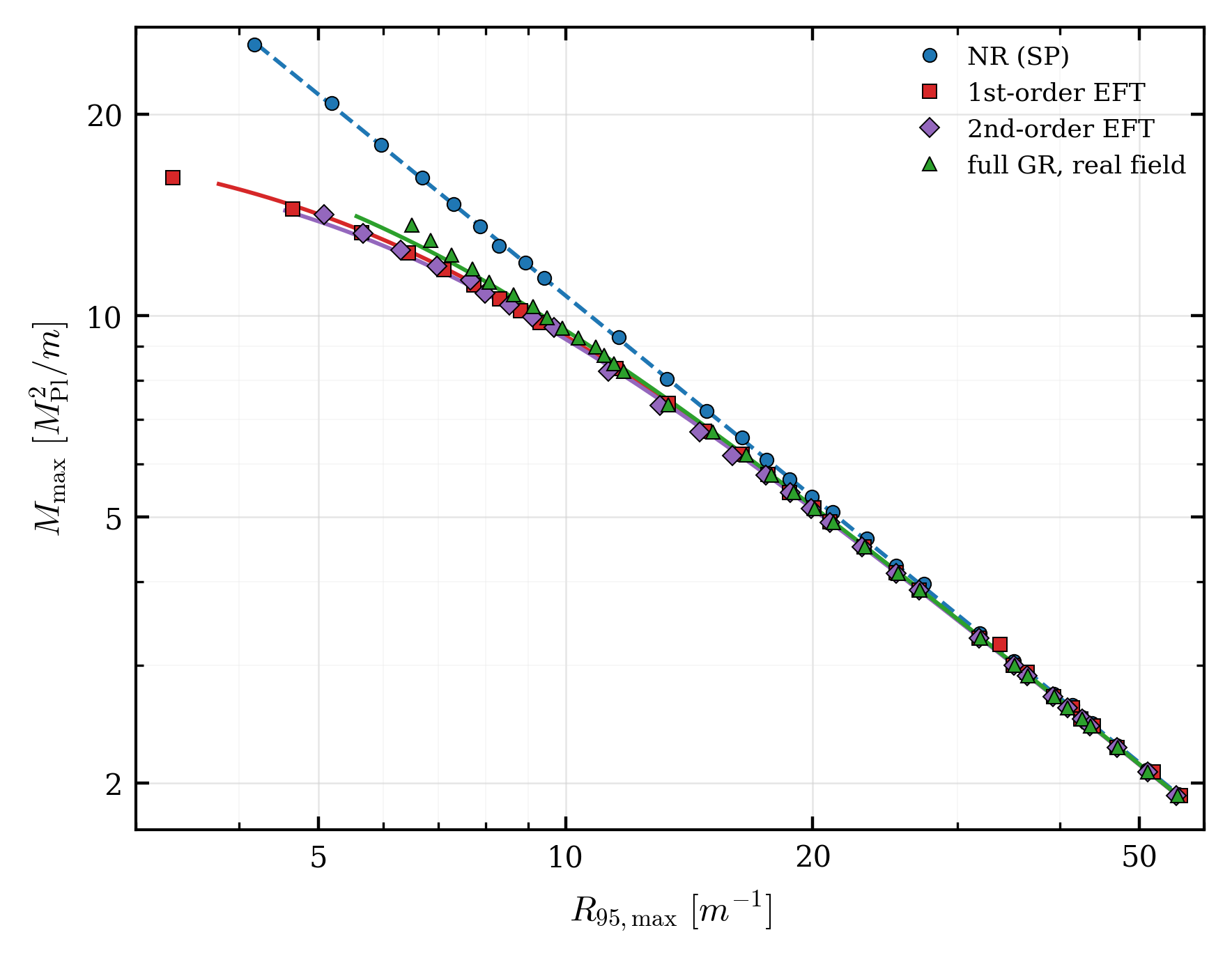}
    \caption{
    Maximum-mass configurations in the mass--radius plane for the different approximation schemes. 
    Each point corresponds to the maximum ADM mass obtained at fixed \(|\tilde{\lambda}|\), with radius defined by \(R_{95}\). 
    The curves are obtained by combining the constrained mass and radius
    fits over \(4\leq|\tilde{\lambda}|\leq700\). 
    }
    \label{fig:mass_radius_max_scaling}
\end{figure}

The maximum-mass sequence in the mass--radius plane is shown in Fig.~\ref{fig:mass_radius_max_scaling}. It therefore summarizes how the critical configuration moves in the \(M_{\max}\)--\(R_{95,\max}\) plane as the attractive self-interaction is varied, and how this behavior changes between the SP, relativistically corrected EFT, and full GR descriptions.

To quantify the accuracy of the approximate descriptions within the common quartic theory without
introducing any dependence on the interpolation fits, we compare the
raw numerical maximum-mass points directly with the full-GR result. {Representative raw turning points
are listed in Appendix~\ref{app:fit}, Table~\ref{tab:raw_max_points}}.
For \(X=\{\mathrm{SP},\mathrm{1st},\mathrm{2nd}\}\), we define
\begin{equation}
    \epsilon_X(|\tilde{\lambda}|)
    =
    100
    \left|
    \frac{M_{\max}^{X}(|\tilde{\lambda}|)}
         {M_{\max}^{\rm GR}(|\tilde{\lambda}|)}
    -1
    \right| .
\end{equation}

\begin{figure*}
    \centering
    \includegraphics[width=0.72\textwidth]{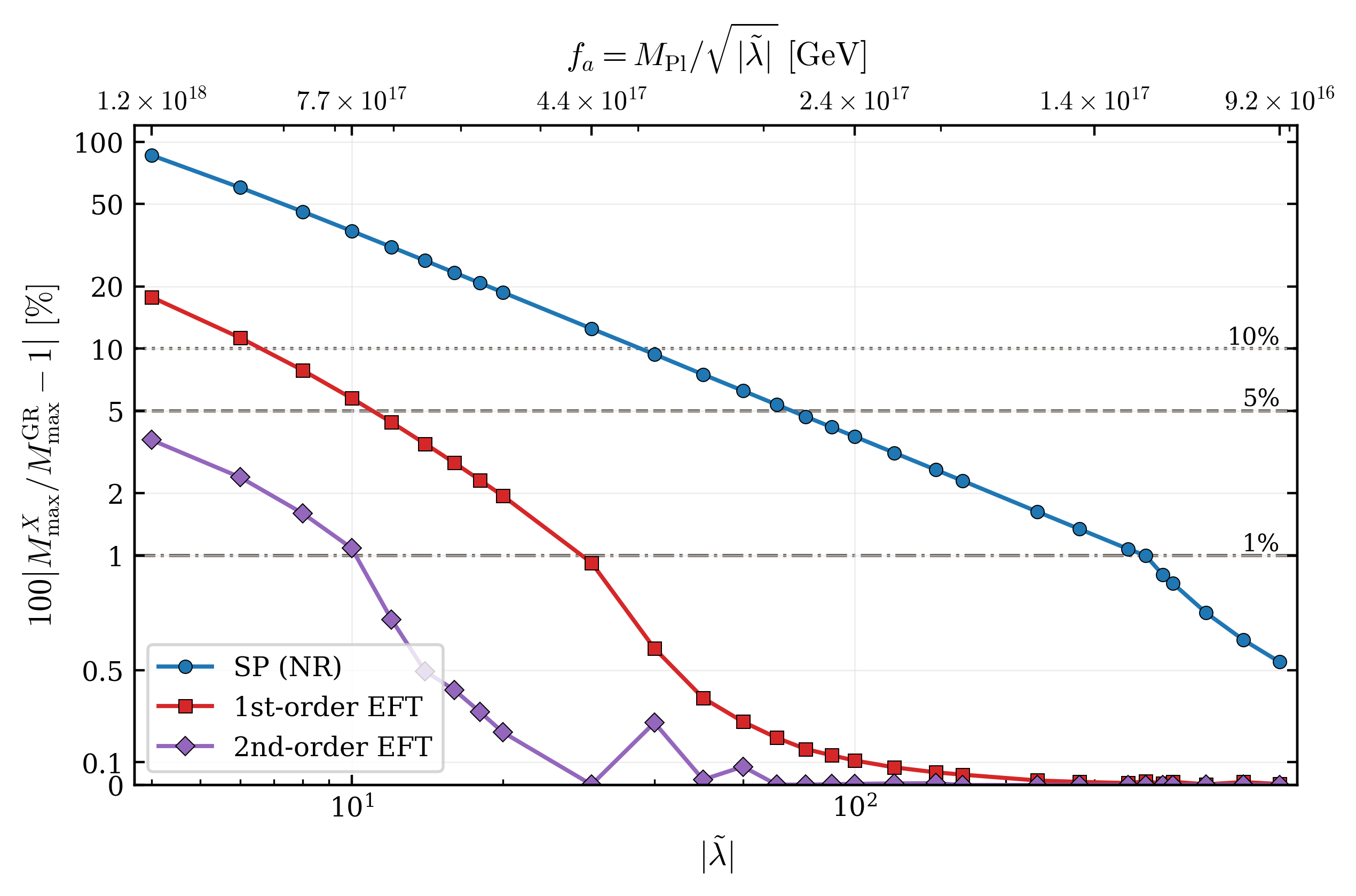}
    \caption{
    Absolute relative difference in the maximum mass predicted by the
    Schr\"odinger--Poisson, first-order EFT, and second-order EFT
    descriptions with respect to the full-GR real scalar result,
    \(\epsilon_X=100|M_{\max}^{X}/M_{\max}^{\rm GR}-1|\).
    The lower axis shows \(|\tilde{\lambda}|\), while the upper
    axis shows the decay constant
    \(f_a=M_{\rm Pl}/\sqrt{|\tilde{\lambda}|}\).
    Horizontal lines indicate the \(1\%\), \(5\%\), and \(10\%\)
    accuracy levels. The comparison uses the raw numerical
    maximum-mass points rather than the interpolation fits.
    }
    \label{fig:accuracy_vs_gr}
\end{figure*}

The comparison is restricted to values of \(|\tilde{\lambda}|\) for
which genuine maximum-mass configurations are available for all four
descriptions. Using 
\begin{equation}
    |\tilde{\lambda}|=\left(\frac{M_{\rm Pl}}{f_a}\right)^2,
\end{equation} the upper
axis of Fig.~\ref{fig:accuracy_vs_gr} shows the corresponding effective
decay constant. The resulting hierarchy is clear. The SP approximation
reaches the \(1\%\) accuracy level at \(|\tilde{\lambda}|\simeq380\),
corresponding to \(f_a\simeq1.3\times10^{17}\,\mathrm{GeV}\); the
first-order EFT reaches it at \(|\tilde{\lambda}|\simeq30\), or
\(f_a\simeq4.5\times10^{17}\,\mathrm{GeV}\), and the second-order EFT
already at \(|\tilde{\lambda}|\simeq10\), or
\(f_a\simeq7.5\times10^{17}\,\mathrm{GeV}\). The relativistic
corrections therefore extend the range of percent-level accuracy in
decay constant by factors of approximately \(3.6\) and
\(6.0\), with most of the gain already achieved at first order.

%%%%%%
%%%%%%%

\subsection{Sensitivity of the quartic axion-potential truncation}
\label{sec:quartic_validity}

The systematic comparison above uses the same attractive quartic
interaction in the SP, relativistic EFT, and full-GR calculations.
For an axion interpretation, however, this interaction represents the
leading nonlinear term in the expansion of the periodic potential
\begin{equation}
    V(\phi)
    =
    m^2 f_a^2
    \left[
        1-\cos\left(\frac{\phi}{f_a}\right)
    \right]
\end{equation}
about its minimum. The quantity controlling this expansion is therefore
the real-field excursion
\begin{equation}
    \max_{t,r}
    \frac{\left|\phi(t,r) \right|}{f_a}
   .
    \label{eq:theta_max}
\end{equation}
Using $|\tilde{\lambda}|=
     \left({M_{\rm Pl}}/{f_a}\right)^2$

we evaluate \(|\phi|/f_a\) at the maximum-mass configuration for
each approximation.

\begin{figure}
    \centering
    \includegraphics[width=0.48\textwidth]{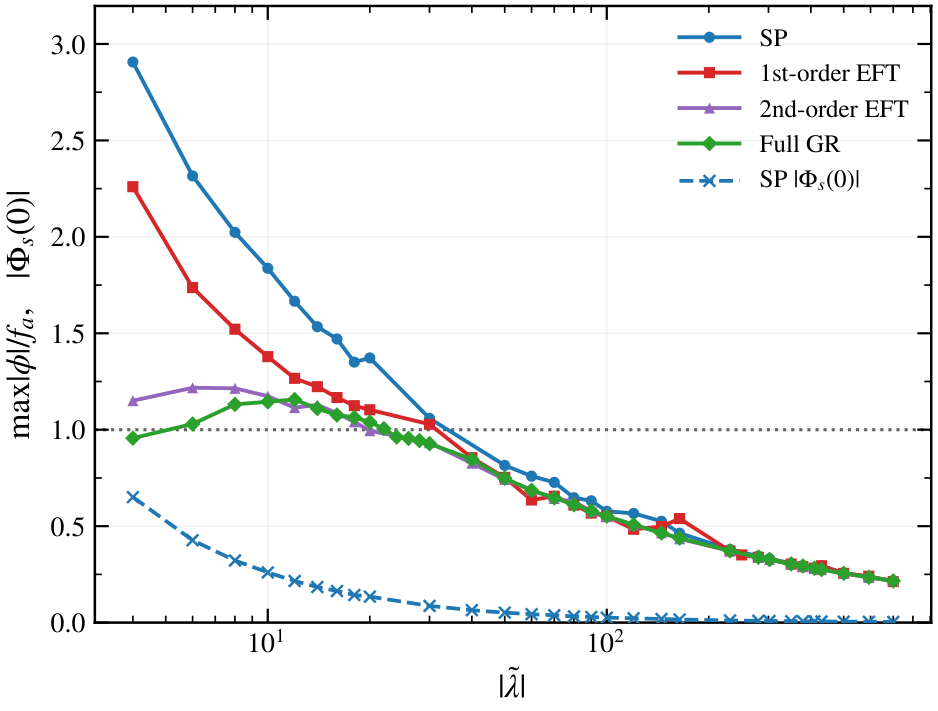}
    \caption{
Maximum real-field excursion,
$\max_{t,r}|\phi|/f_a$, evaluated at the maximum-mass
configuration as a function of the self-interaction parameter
$|\tilde{\lambda}|$ for the SP, first- and second-order EFT,
and full-GR descriptions. The horizontal dotted line indicates
$\max|\phi|/f_a=1$ as a reference for order-unity field
excursions. In the weak to moderate coupling regime, some
maximum-mass configurations reach
$\max|\phi|/f_a=\mathcal{O}(1)$, motivating an explicit
comparison between the quartic truncation and the complete
cosine potential. The dashed curve shows the maximum gravitational potential at the center of the Schr\"odinger--Poisson maximum-mass
configuration. It reaches $|\Phi_s (0)|\simeq 0.65$ at
$|\tilde{\lambda}|=4$, whereas the Newtonian approximation
requires $|\Phi_s|\ll1$.
}
    \label{fig:phi_over_fa}
\end{figure}

As shown in Fig.~\ref{fig:phi_over_fa}, the large-
\(|\tilde{\lambda}|\) configurations satisfy
\(|\phi|/f_a\ll1\), while in the weak to moderate coupling regime
some of the maximum-mass solutions reach
\(|\phi|/f_a=\mathcal{O}(1)\).
The latter is precisely the region in which the quartic expansion is
no longer parametrically guaranteed.
Order-unity field excursions should not be interpreted as a sharp
failure boundary. They instead identify the part of parameter space
where the higher terms of the periodic potential must be tested
explicitly. The complete cosine potential has also been used in fully relativistic
axion-star simulations. In ~\cite{Helfer:2016ljl}, the authors evolved the full nonlinear
Einstein equations with the complete cosine potential and identified
long-lived oscillating configurations, black-hole formation, and
dispersal.
Related full-cosine relativistic calculations were performed by ~\cite{Michel:2018, Widdicombe:2018}.
These works provide useful dynamical context, although their phase
boundaries correspond to evolved initial data families rather than to
the equilibrium turning points considered here.
\begin{figure}
    \centering
    \includegraphics[width=0.5\textwidth]{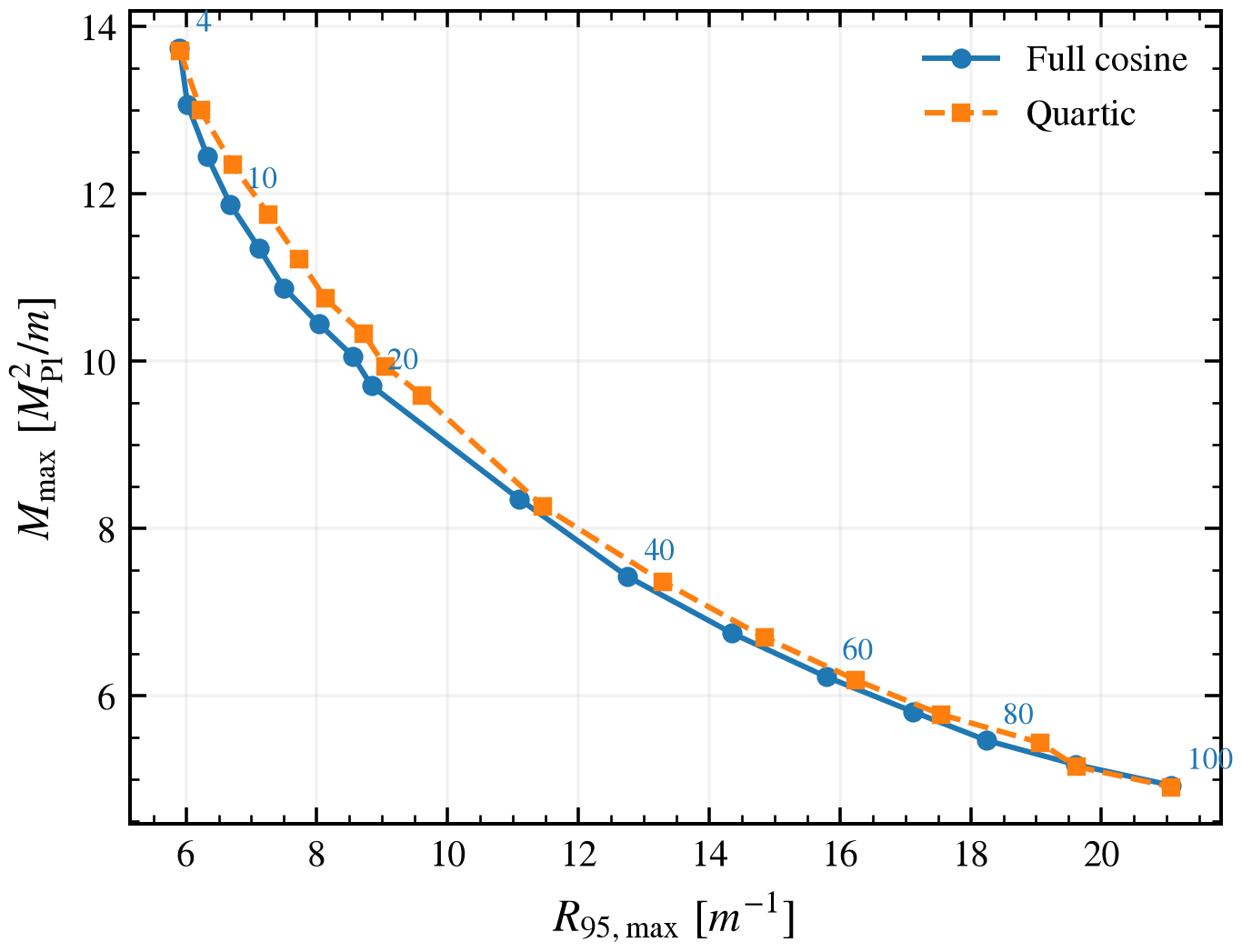}
    \caption{
Maximum-mass configurations in the
\(M_{\max}\)--\(R_{95,\max}\) plane obtained using the complete
single-cosine potential and its quartic truncation. Each point
corresponds to the maximum-mass configuration at fixed
\(|\tilde{\lambda}|\). Both calculations use the same
full-GR periodic equilibrium construction. The close agreement
provides a direct test of the quartic-potential truncation in the
region where \(\max|\phi|/f_a\) approaches unity.}
    \label{fig:cosine_quartic_mr}
\end{figure}
For a direct test of the potential truncation, we solve the same
periodic full-GR boundary-value problem using both the quartic and
complete cosine potentials. Fig.~\ref{fig:cosine_quartic_mr}
compares the corresponding maximum-mass configurations in the
\(M_{\max}\)--\(R_{95,\max}\) plane for
\(4\leq|\widetilde{\lambda}|\leq100\).
\begin{figure*}
    \centering
    \includegraphics[width=1.0\textwidth]{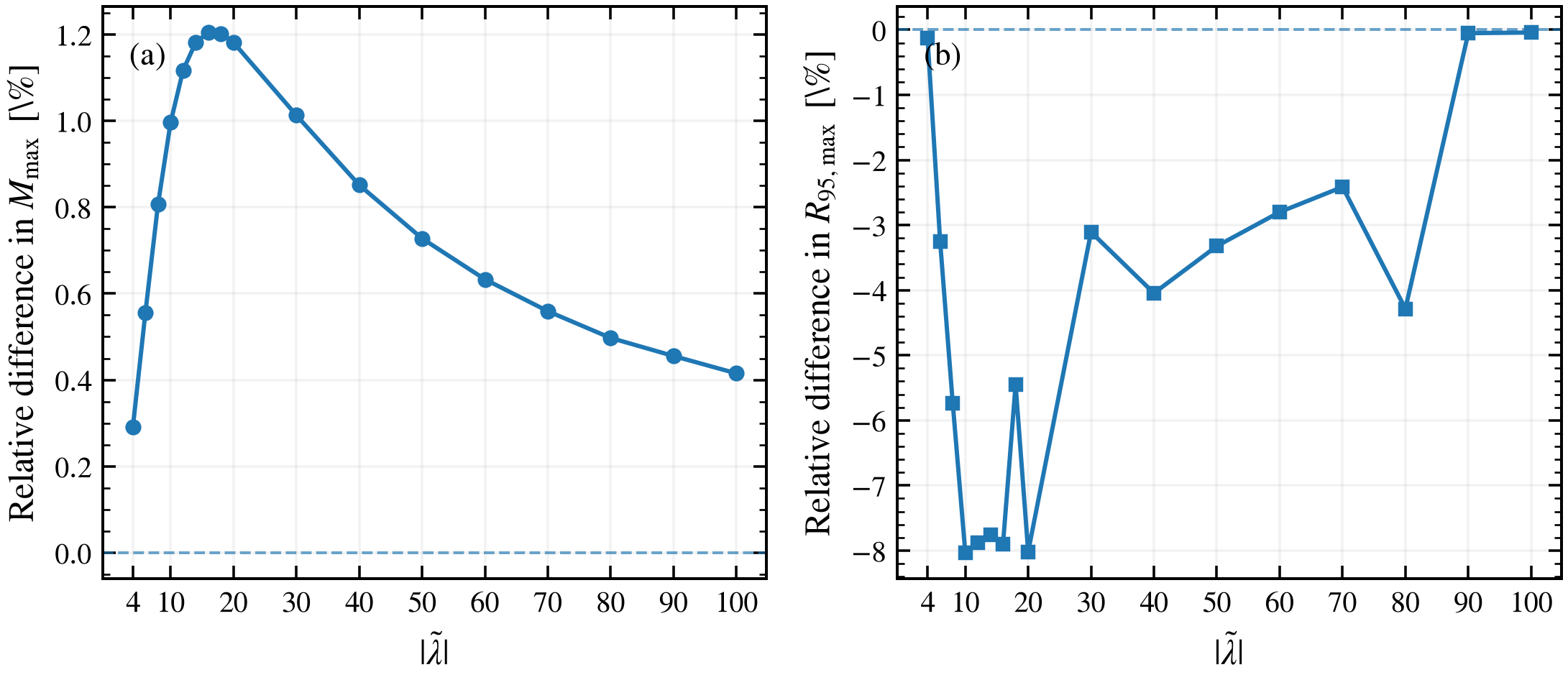}
    \caption{Relative differences between the maximum-mass configurations obtained
with the complete cosine potential and its quartic truncation as a
function of $|\tilde{\lambda}|$. Panel (a) shows the relative
difference in $M_{\max}$ and panel (b) the corresponding difference in
$R_{95,\max}$, with the quartic result used as the reference:
$\Delta X=100(X^{\rm cos}-X^{\rm quart})/X^{\rm quart}$.
Positive (negative) values therefore indicate that the full-cosine
result is larger (smaller) than the quartic result.
}
    \label{fig:cosine_quartic_relative}
\end{figure*}
The two sequences remain close over the entire interval. The maximum
mass is only weakly affected by restoring the higher-order terms of the
periodic potential, whereas the enclosed-mass radius displays a
somewhat larger sensitivity. The size of the correction is shown more directly in
Fig.~\ref{fig:cosine_quartic_relative}. Over the range tested, the
change in \(M_{\max}\) remains at the percent level, while the
correction to \(R_{95,\max}\) reaches several percent. Thus the
radius is more sensitive than the maximum mass to the higher-order
terms in the axion potential.

Two previous studies clarify how our result fits into the broader landscape. Ref.~\cite{Eby2019} worked in the QCD regime, $f_a\sim 10^{11}$~GeV (large $|\tilde{\lambda}|$ in our convention), and showed that even after resumming the axion self-interaction into a Bessel-function Gross--Pitaevskii--Poisson description, the nonrelativistic prediction departs from the relativistic Einstein--Klein--Gordon result --- evaluated within the Ruffini--Bonazzola (RB) bound-state ansatz \cite{ruffini_1969} --- once the binding becomes strong. In their case, strong binding is reached only \emph{beyond} the first dilute maximum, in the post-maximum transition toward the dense regime. Ref.~\cite{Eby2021} instead studied nearly Planck-scale $f_a$ (small $|\tilde{\lambda}|$) using a weak-gravity expansion with leading relativistic bound-state corrections. In their notation, the binding is measured by $\Delta=\sqrt{1-\epsilon^2}$, where $\epsilon\, m$ is the single-particle bound-state energy, and the nonrelativistic dilute branch requires
\begin{equation}
    \Delta \ll 1.
    \label{rb_scale_nr}
\end{equation}
At the conventional nonrelativistic maximum they find
\begin{equation}
    \Delta_c\simeq
    \frac{1}{\sqrt{\kappa_c|\tilde{\lambda}|}},
    \label{eq:binding-turning-point}
\end{equation}
with $\kappa_c\simeq0.34$ the critical effective coupling of the EKG--RB ansatz ($\kappa \geq \kappa_c$ for stable solutions). Decreasing $|\tilde{\lambda}|$ therefore drives the first maximum itself toward stronger binding: $\Delta_c\simeq0.54$ at $|\tilde{\lambda}|=10$, already violating \eqref{rb_scale_nr}, and would reach approximately $0.86$ at $|\tilde{\lambda}|=4$, outside the controlled range of their expansion. The two works probe complementary manifestations of the same limitation: at small $f_a$, strong binding appears past the dilute maximum; at large $f_a$, it appears at the first maximum itself.

Our analysis lies in the second regime: the first turning point is already relativistic at low $|\tilde{\lambda}|$, and dominates the SP--GR discrepancy. The direct evidence is that the first- and second-order EFT corrections substantially reduce the gap while retaining exactly the same quartic interaction, demonstrating that a significant part of the discrepancy is structural rather than perturbative in the expansion of the potential. In addition, the full-GR maximum mass changes only at the percent level when the quartic truncation is replaced by the matched single-cosine potential. Resumming the potential controls the $\phi/f_a$ expansion, but does not resum the independent relativistic corrections associated with finite binding energy, relativistic gradients, and gravitational backreaction. We cannot rule out, however, that the anharmonic correction to SP is larger than what we measure in full GR;  as Fig.~\ref{fig:phi_over_fa} shows, the SP maximum-mass configurations reach substantially larger field excursions at the smallest $|\tilde{\lambda}|$, so a dedicated all-orders SP calculation with the matched cosine potential would be needed to isolate quantitatively the residual anharmonic contribution.
The same figure shows the central gravitational potential of the SP
maximum-mass configuration, which reaches \(|\Phi_s(0)|\simeq 0.65\) at
\(|\tilde{\lambda}|=4\). Both \(\epsilon_\lambda\) and \(\epsilon_g\) are no longer small. This is the origin of the SP--GR
discrepancy: the SP solution predicts a configuration too compact for
the approximation that produced it. The same breakdown explains why
\(R_{95,\max}\) is nonmonotonic between the first- and second-order
EFT systems, since an asymptotic series with \(\epsilon=\mathcal{O}(1)\)
need not improve term by term.

Two scope limitations are worth stating explicitly. First, our analysis terminates at the first turning point and does not follow the post-maximum unstable branch. Second, the purely attractive quartic theory used in our common comparison does not by itself admit a physically stabilized dense branch; references to a transition or nominal dense regime elsewhere in the literature pertain to all-orders axion models. In summary, the large low-$|\tilde{\lambda}|$ discrepancy along our first-maximum sequence reflects the turning point itself becoming strongly bound and relativistic. At larger $|\tilde{\lambda}|$, the first maximum remains weakly bound and is adequately described by SP.

\section{Discussion and Summary}
\label{sec:discussion}

We have compared the maximum-mass configurations of axion stars in four descriptions: the Newtonian Schr\"odinger--Poisson approximation, the first- and second-order relativistically corrected EFTs of Ref.~\cite{salehian_beyond_2021}, and the full-GR real-scalar oscillaton. 
Using a common quartic interaction and consistent definitions of $M_{\max}$ and $R_{95}$, all four descriptions recover the common large-$|\tilde{\lambda}|$ dilute scaling $M_{\max}\propto|\tilde{\lambda}|^{-1/2}$  expected from the nonrelativistic attractive-self-interaction limit~\cite{salehian_beyond_2021}, while their finite-coupling predictions separate as the first turning point becomes strongly bound. Our main findings are:

\begin{itemize}
    \item We map the first maximum-mass turning point over $4\leq|\tilde{\lambda}|\leq 700$, corresponding to $f_a\gtrsim9\times10^{16}\,{\rm GeV}$ in the SP, first- and second-order EFT, and full-GR descriptions, using common mass and radius prescriptions, and test the temporal-harmonic and potential truncations directly in full GR.
    
    \item Using full GR as reference, SP overestimates $M_{\max}$ by approximately $86\%$, $37\%$, $19\%$, and $9.4\%$ at $|\tilde{\lambda}|=4,10,20,40$, and underestimates $R_{95}$ by $\sim 30\%$ at $|\tilde{\lambda}|=4$.
    
    \item The relativistic EFTs substantially reduce the mass discrepancy while retaining the same quartic interaction: the $1\%$ accuracy threshold in $M_{\max}$ moves from $|\tilde{\lambda}|\gtrsim379$ (SP) to $\gtrsim30$ (first order) and $\gtrsim10.5$ (second order), see Fig.~\ref{fig:accuracy_vs_gr}. The improvement need not be strictly monotonic order by order; the radius, in particular, is nonmonotonic between first and second order.
    
    \item The large low-$|\tilde{\lambda}|$ discrepancy is driven by increasing binding rather than by the coupling strength alone: decreasing $|\tilde{\lambda}|$ (equivalently, increasing $f_a$) moves the first turning point into a regime where binding, gradients, and gravitational backreaction are no longer uniformly nonrelativistic~\cite{Eby2019,Eby2021}.
    
    \item In full GR, enlarging the temporal-harmonic basis stabilizes the structural observables rapidly, and replacing the quartic truncation by the matched single-cosine potential shifts $M_{\max}$ only at the percent level over $4\leq|\tilde{\lambda}|\leq100$, with a somewhat larger effect ($\sim$~several percent) in $R_{95}$. These tests probe the potential and harmonic truncations, not dynamical stability or lifetime.
\end{itemize}

The principal new result of this work is not the existence of the EFT hierarchy of Ref.~\cite{salehian_beyond_2021}, but the coupling-wide accuracy map of the first turning point supplemented by direct full-GR tests of the harmonic and potential truncations. The physical origin of the SP--GR gap and its interpretation in light of \cite{Eby2019,Eby2021} were analyzed in Sec.~\ref{sec:quartic_validity}: the EFT hierarchy moves the quartic prediction toward full GR without changing the potential, isolating a genuine relativistic contribution to the discrepancy that resummation of the axion potential alone would not eliminate. A dedicated all-orders SP calculation with the matched cosine potential would nevertheless be required to isolate quantitatively the residual anharmonic contribution to SP at the smallest $|\tilde{\lambda}|$.

\subsection*{Phenomenological implications}

Many applications of axion stars use $M_{\max}$, $R_{95}$, the
compactness, or the density profile as structural
inputs~\cite{Helfer:2016ljl,Chavanis:2017loo,Hertzberg:2018zte,Hertzberg:2020dbk,Escudero:2023vgv},
so a nonrelativistic estimate of $M_{\max}$ shifts the threshold that
separates subcritical from supercritical configurations. The
following examples illustrate where this matters.

\paragraph{Axion-star explosions.} 
In the scenario of Refs.~\cite{Du:2023jxh,Escudero:2023vgv}, constraints from the CMB optical depth and projected $21\,{\rm cm}$ sensitivity depend on the abundance of axion stars that become supercritical and decay into photons. Since the critical soliton mass enters the host-halo threshold through the core--halo relation $M_s\propto M_h^\alpha$, a fractional change in the critical soliton mass, $M_{s,\rm crit}\to q\, M_{s,\rm crit}$, maps to
\begin{equation}
    M_{h,\rm crit}\to q^{1/\alpha}\, M_{h,\rm crit}.
    \label{eq:halo-threshold-shift}
\end{equation}
For the benchmark $\alpha=2/5$ of Ref.~\cite{Escudero:2023vgv} and our raw equilibrium ratio $q=M_{\max}^{\rm GR}/M_{\max}^{\rm SP}\simeq 0.54$ at $|\tilde{\lambda}|=4$, Eq.~\eqref{eq:halo-threshold-shift} gives $q^{1/\alpha}\simeq0.21$, a $79\%$ reduction of the inferred host-halo threshold; at $|\tilde{\lambda}|=20$, where $q\simeq0.84$, the factor is $\simeq 0.65$. This is a scaling exercise, not a revised bound: the event rate also depends on the halo mass function, merger history, scatter in the core--halo relation, and competition among instability channels.

\paragraph{Electromagnetic instability.} 
Fully nonlinear $3+1$ simulations of compact axion stars coupled to photons through $\mathcal{L}\supset g_{a\gamma}\phi F^{\mu\nu}\tilde{F}_{\mu\nu}$ show that they become unstable above a critical coupling scaling as $g_{a\gamma}^{\rm crit}\propto M_s^{-1.35}$, with peak emission at $\omega\sim 1/R_s$~\cite{Chung-Jukko:2023wlc}. Applying these scalings illustratively, a mass ratio $q_M\simeq 0.54$ would shift the critical coupling by $q_M^{-1.35}\simeq 2.3$, and a radius change by $q_R$ would shift the emission frequency by $q_R^{-1}$. These mass- and radius-only rescalings neglect correlated changes in compactness and field amplitude; a quantitative prediction requires evolving the coupled axion--electromagnetic system on the corrected profiles.

\paragraph{Microlensing, mergers, and resonance.} 
Microlensing bounds include finite-lens and finite-source effects that
depend on the projected mass distribution relative to the Einstein
radius~\cite{Fujikura:2021omw}, and merger-triggered photon resonance
depends on the density, size, and coherence of the
configuration~\cite{Hertzberg:2018zte,Hertzberg:2020dbk}. Both are
therefore sensitive to the relativistic changes in $R_{95}$ and in the
density profile, most strongly at low $|\tilde{\lambda}|$; the sign
and size of the effect require repeating each calculation with the
corrected profiles.
\subsection*{Scope and caveats}

Our comparison establishes equilibrium structure; it does not determine dynamical outcomes. Numerical-relativity simulations have shown that axion-star collapse can yield long-lived configurations, dispersal via gravitational cooling, or black-hole formation~\cite{Helfer:2016ljl,Michel:2018,Widdicombe:2018}, and that collisions with black holes and neutron stars require full $3+1$ evolution~\cite{Clough:2018exo}. Our results provide the corrected structural inputs from which such dynamics would begin.

A separate limitation concerns the lifetime of the real-scalar configurations. The Fourier construction determines the localized periodic core and its structural observables, but exactly periodic real massive scalar solutions generically carry a small radiative tail, and physical finite-mass oscillatons slowly lose energy~\cite{Page2004,Grandclement2011}; strong anharmonicity and number-changing processes can further shorten the lifetime in axion models~\cite{Visinelli:2017ooc,Eby2019,Eby2021}. A dedicated time-dependent evolution with radiative boundary conditions would instead be required to determine the outgoing flux.

\subsection*{Summary}

The nonrelativistic Schr\"odinger--Poisson approximation reliably
captures the large-$|\tilde{\lambda}|$ scaling of the maximum-mass
branch, but significantly overestimates $M_{\max}$ in the weak-to-moderate coupling regime, corresponding to decay constants
$f_a\gtrsim 9\times 10^{16}\,{\rm GeV}$.  For the smaller values characteristic
of the QCD axion, all four descriptions converge to the common dilute
scaling. The weak-coupling regime is precisely where phenomenological
predictions that depend on the critical soliton mass, radius, or
compactness should be treated with care. Our comparison identifies
where nonrelativistic structural predictions remain adequate and where
relativistically corrected or fully general-relativistic calculations
are required.

\begin{acknowledgments}
E.D.S. acknowledges support from FONDECYT Project N° 1251141 (Agencia Nacional de Investigaci\'on y Desarrollo, Chile). E.D.S. is grateful to Prof. Shigeki Matsumoto of the Kavli Institute for the Physics and Mathematics of the Universe (Kavli IPMU), The University of Tokyo, for his hospitality during a research stay at IPMU supported by this project, where this work was finalized. P.A. acknowledges support
from FONDECYT project N° 1251613.
 \end{acknowledgments}

\appendix
\begin{center}
    \textbf{Appendices}
\end{center}

\section{Numerical Method and Implementation}
\label{app:numerics}

In this appendix we describe the numerical procedures used for the three systems considered in this work: the first-order corrected EFT system, the second-order corrected EFT system, and the full Einstein--Klein--Gordon (EKG) system. In all cases, the goal is to construct the nodeless ground-state branch and extract the corresponding mass--radius relation.

% ============================================================
\subsection{First-order corrected EFT system}
\label{app:numerics_first_order}

We first solve the first-order relativistically corrected Schrödinger--Poisson system derived in Sec.~\ref{subsec:first-order} and written explicitly in Eqs.~\eqref{eq_f_adim}--\eqref{eq_phi_adim}. After the rescaling introduced in the main text, the equations form a coupled nonlinear system of ordinary differential equations,
\begin{equation}
	\mathcal{E}_f[f,\Phi_s;\tilde{\mu},\tilde{\lambda}]=0,
	\qquad
	\mathcal{E}_{\Phi}[f,\Phi_s;\tilde{\mu},\tilde{\lambda}]=0.
	\label{eq:app_first_order_symbolic_system}
\end{equation}
Here \(f\equiv \tilde f\), and the radial coordinate is the dimensionless radius defined in Sec.~\ref{subsec:dimensionless_parameters}. For fixed \((\tilde\mu,\tilde\lambda)\), the system is solved as a two-parameter shooting problem.

Spherical symmetry and regularity at the origin require
\begin{equation}
	f'(0)=0,
	\qquad
	\Phi_s'(0)=0.
	\label{eq:app_first_order_regular_bc}
\end{equation}
To avoid the coordinate singularity at \(r=0\), we begin the integration at a small radius \(r=\Delta\ll1\). The initial values are obtained from the Taylor expansions
\begin{align}
	f(r)
	&=
	\alpha_0+\alpha_2 r^2+\alpha_4 r^4+\mathcal{O}(r^6),
	\label{eq:app_first_order_f_taylor}
	\\
	\Phi_s(r)
	&=
	\beta_0+\beta_2 r^2+\beta_4 r^4+\mathcal{O}(r^6).
	\label{eq:app_first_order_phi_taylor}
\end{align}
The free shooting parameters are
\begin{equation}
	\alpha_0=f(0),
	\qquad
	\beta_0=\Phi_s(0).
	\label{eq:app_first_order_free_params}
\end{equation}
The coefficients \(\alpha_2,\beta_2,\alpha_4,\beta_4\) are determined algebraically by substituting Eqs.~\eqref{eq:app_first_order_f_taylor}--\eqref{eq:app_first_order_phi_taylor} into Eqs.~\eqref{eq_f_adim}--\eqref{eq_phi_adim} and solving order by order in \(r\). These Taylor-expanded expressions are used to initialize the numerical integration at \(r=\Delta\).

At large radius, physical solutions must be localized. The scalar profile satisfies
\begin{equation}
	f(r)\xrightarrow{r\to\infty}0,
	\label{eq:app_first_order_f_localized}
\end{equation}
while the gravitational potential approaches
\begin{equation}
	\Phi_s(r)\xrightarrow{r\to\infty}\Phi_\infty-\frac{C}{r}.
	\label{eq:app_first_order_phi_asymptotic}
\end{equation}
For the convention used in Eqs.~\eqref{eq_f_adim}--\eqref{eq_phi_adim}, the asymptotic normalization is chosen so that \(\Phi_\infty=0\).

At large \(r\), the scalar equation linearizes to
\begin{equation}
	f''+\frac{2}{r}f'-\Omega^2 f=0,
	\qquad
	\Omega(\tilde\mu)=\sqrt{2\tilde\mu-\tilde\mu^2}.
	\label{eq:app_first_order_linear_tail}
\end{equation}
The general asymptotic solution is
\begin{equation}
	f_\infty(r)
	=
	\frac{a_1}{r}e^{-\Omega r}
	+
	\frac{a_2}{2\Omega r}e^{+\Omega r}.
	\label{eq:app_first_order_f_asymptotic}
\end{equation}
Normalizability requires \(a_2=0\). The gravitational potential has the asymptotic form
\begin{equation}
	\Phi_s(r)=b_2-\frac{b_1}{r}.
	\label{eq:app_first_order_phi_asymptotic_coeffs}
\end{equation}
At a finite matching radius \(r=r_\infty\), we evaluate
\begin{align}
	a_1(r,\tilde\mu)
	&=
	\frac{e^{+\Omega r}}{2\Omega}
	\left[
	f(r)(-1+r\Omega)-r f'(r)
	\right],
	\label{eq:app_first_order_a1}
	\\
	a_2(r,\tilde\mu)
	&=
	e^{-\Omega r}
	\left[
	f(r)(1+r\Omega)+r f'(r)
	\right],
	\label{eq:app_first_order_a2}
	\\
	b_1(r)
	&=
	r^2\Phi_s'(r),
	\label{eq:app_first_order_b1}
	\\
	b_2(r)
	&=
	\Phi_s(r)+r\Phi_s'(r).
	\label{eq:app_first_order_b2}
\end{align}
The shooting conditions are then
\begin{equation}
	a_2(r_\infty,\tilde\mu)=0,
	\qquad
	b_2(r_\infty)=0.
	\label{eq:app_first_order_matching_conditions}
\end{equation}
The first condition removes the growing scalar mode, and the second fixes the asymptotic normalization of the gravitational potential.

For each fixed pair \((\tilde\mu,\tilde\lambda)\), the numerical procedure is:
\begin{enumerate}
	\item Choose an initial guess for \((\alpha_0,\beta_0)\).
	\item Initialize the fields at \(r=\Delta\) using the Taylor expansions.
	\item Integrate the coupled equations outward to \(r=r_\infty\).
	\item Evaluate \(a_2(r_\infty,\tilde\mu)\) and \(b_2(r_\infty)\).
	\item Adjust \((\alpha_0,\beta_0)\) until the two matching conditions in Eq.~\eqref{eq:app_first_order_matching_conditions} are satisfied.
\end{enumerate}
The matching radius is increased until the scalar tail is sufficiently resolved. Solutions are rejected if the scalar field develops a radial node or if the numerical integration becomes singular before reaching the matching radius.

The soliton mass is computed as
\begin{equation}
	M_s
	=
	4\pi
	\int_0^{r_{\max}}dr\,r^2
	\left[
	f^2
	\left(
	1-\frac{7}{2}\Phi_s
	\right)
	+
	\frac{1}{2}(f')^2
	+
	\frac{\tilde{\lambda}}{16}f^4
	\right].
	\label{soliton_mass}
\end{equation}
The effective radius \(R_{95}\) is defined by
\begin{equation}
	M(r<R_{95})=0.95M_s.
	\label{eq:app_first_order_R95}
\end{equation}

To construct the mass--radius relation at fixed \(\tilde\lambda\), the updated implementation scans over a sequence of \(\tilde\mu\) values. The optimized values of \((\alpha_0,\beta_0)\) obtained for one value of \(\tilde\mu\) are used as the initial guess for the next value. This continuation procedure improves convergence and keeps the solver on the same nodeless branch.

The maximum mass is determined in two stages. First, a coarse scan over \(\tilde\mu\) is used to locate the approximate maximum of the sequence. If the largest mass in the coarse scan occurs near \(\tilde\mu_i\), a second scan is performed in a smaller interval around that point, using the neighboring values \(\tilde\mu_{i-1}\) and \(\tilde\mu_{i+1}\) as the boundaries of the refined interval. The solution at the coarse maximum is used as the initial seed for the refined scan. The final maximum-mass point is therefore obtained from the local refined scan rather than directly from the coarse grid.

% ============================================================
\subsection{Second-order corrected EFT system}
\label{app:numerics_second_order}

The second-order corrected EFT system is solved using the same shooting philosophy, but with a larger set of radial functions. In the notation used in the numerical implementation,
\begin{equation}
	F\equiv \tilde f,
	\quad
	\Phi\equiv \widehat{\Phi}_s,
	\quad
	\Psi\equiv \Psi_s,
	\quad
	A\equiv \Phi^{(1)}_2,
	\quad
	B\equiv \Psi^{(2)}_2.
	\label{eq:app_second_order_notation}
\end{equation}
The bulk equations are those given in Eqs.~\eqref{eq_f_second_order}--\eqref{eq_phi2_constraint}. After the field redefinition in Eq.~\eqref{eq:mu_removed_redefinition}, the chemical-potential parameter does not appear explicitly in the differential equations. Instead, it fixes the asymptotic value of the shifted lapse potential,
\begin{equation}
	\Phi_\infty(\tilde\mu)
	=
	\tilde\mu
	+
	\frac{1}{2}\tilde\mu^2
	+
	\frac{1}{3}\tilde\mu^3.
	\label{eq:app_second_order_phi_infty}
\end{equation}

As before, the integration starts at \(r=\Delta\ll1\). Regularity at the origin implies that the radial derivatives vanish there. We write
\begin{align}
	F(r)&=F_0+F_2r^2+\mathcal{O}(r^4),
	\label{eq:app_second_order_F_taylor}
	\\
	\Phi(r)&=\Phi_0+\Phi_2r^2+\mathcal{O}(r^4),
	\label{eq:app_second_order_Phi_taylor}
	\\
	\Psi(r)&=\Psi_0+\Psi_2r^2+\mathcal{O}(r^4),
	\label{eq:app_second_order_Psi_taylor}
	\\
	A(r)&=A_0+A_2r^2+\mathcal{O}(r^4),
	\label{eq:app_second_order_A_taylor}
	\\
	B(r)&=B_0+B_2r^2+\mathcal{O}(r^4).
	\label{eq:app_second_order_B_taylor}
\end{align}
The coefficients \(F_2,\Phi_2,\Psi_2,A_2,B_2\) are obtained by substituting these expansions into the second-order equations. The shooting parameters are therefore
\begin{equation}
	F_0,
	\qquad
	\Phi_0,
	\qquad
	\Psi_0,
	\qquad
	A_0,
	\qquad
	B_0.
	\label{eq:app_second_order_shooting_params}
\end{equation}

At large radius, the scalar profile again has an exponentially decaying tail. In the second-order system, the decay scale is obtained from the linearized equation around the shifted asymptotic potential:
\begin{equation}
	\Omega^2
	=
	2\Phi_\infty
	\left(
	1-\Phi_\infty+\frac{2}{3}\Phi_\infty^2
	\right).
	\label{eq:app_second_order_omega}
\end{equation}
For small \(\tilde\mu\), this reduces to
\begin{equation}
	\Omega^2=2\tilde\mu-\tilde\mu^2+\mathcal{O}(\tilde\mu^3),
	\label{eq:app_second_order_omega_expansion}
\end{equation}
which agrees with the first-order EFT tail at the corresponding order.

The boundary conditions imposed at \(r=r_\infty\) are
\begin{align}
	F+\frac{rF'}{1+\Omega r}&=0,
	\label{eq:app_second_order_F_bc}
	\\
	\Phi-\Phi_\infty+r\Phi'&=0,
	\label{eq:app_second_order_Phi_bc}
	\\
	\Psi+r\Psi'&=0,
	\label{eq:app_second_order_Psi_bc}
	\\
	A+rA'&=0,
	\label{eq:app_second_order_A_bc}
	\\
	B&=0.
	\label{eq:app_second_order_B_bc}
\end{align}
These conditions impose a decaying scalar field, \(1/r\) asymptotic behavior for the metric potentials, and a vanishing asymptotic value for the nonzero mode \(B\). The condition \(B\to0\) is required because a nonzero constant contribution would source an unphysical asymptotic behavior in the equation for \(A\).

The five shooting parameters in Eq.~\eqref{eq:app_second_order_shooting_params} are adjusted until the five residuals in Eqs.~\eqref{eq:app_second_order_F_bc}--\eqref{eq:app_second_order_B_bc} vanish. The integration radius is increased when the scalar tail has not sufficiently decayed. Candidate solutions are rejected if the scalar profile develops nodes or if any of the fields grows beyond the numerical tolerance.

The second-order mass is computed using Eq.~\eqref{eq_mass_second_order}. In the code notation of Eq.~\eqref{eq:app_second_order_notation}, this becomes
\begin{align}
	M_s
	&=
	4\pi\int_0^{r_{\max}}dr\,r^2
	[
	F^2
	(
	1-\Phi-\frac{5}{2}\Psi
	+\frac{25}{8}\Psi^2
	+\frac{5}{2}\Phi\Psi\nonumber\\
	&
	+\Phi^2
	+A
	-\frac{3F^2}{32}
	)
	+
	\frac{1}{2}(F')^2
	\left(
	1-\frac{1}{2}\Psi
	\right)\nonumber\\
	&
	+
	\frac{\tilde\lambda F^4}{16}
	\left(
	1-\frac{5}{2}\Psi
	\right)
	+
	\frac{13\tilde\lambda^2F^6}{9216}
	].
	\label{eq:app_second_order_mass}
\end{align}
The radius \(R_{95}\) is again determined from
\begin{equation}
	M(r<R_{95})=0.95M_s.
	\label{eq:app_second_order_R95}
\end{equation}

The scan over \(\tilde\mu\) is performed by continuation, as in the first-order calculation. The optimized central values
\begin{equation}
	(F_0,\Phi_0,\Psi_0,A_0,B_0)
	\label{eq:app_second_order_seed}
\end{equation}
are used as the initial seed for the next value of \(\tilde\mu\). The maximum of the second-order mass sequence is then refined by repeating the scan in a smaller interval around the preliminary maximum.

% ============================================================
\subsection{Full EKG system}
\label{app:numerics_full_GR}

For comparison with the EFT results, we also solve the full real-field EKG system in spherical symmetry. Unlike the EFT calculation, the full-GR system does not average over the fast scalar oscillations. Instead, the real scalar field and the geometry are solved as periodic functions of time.

The full-GR calculation uses the areal radius metric
\begin{equation}
	ds^2
	=
	-\alpha^2(t,x)dt^2
	+
	a^2(t,x)dx^2
	+
	x^2d\Omega^2,
	\label{eq:app_full_GR_metric}
\end{equation}
where \(x\) is the dimensionless areal radius. The quartic axion potential is
\begin{equation}
	U(\phi)
	=
	\frac{1}{2}\phi^2
	+
	\frac{\tilde\lambda}{24}\phi^4,
	\qquad
	\frac{dU}{d\phi}
	=
	\phi
	+
	\frac{\tilde\lambda}{6}\phi^3.
	\label{eq:app_full_GR_potential}
\end{equation}

Following the standard oscillaton construction, we define
\begin{equation}
	A(t,x)=a^2(t,x),
	\qquad
	C(t,x)=\left(\frac{a(t,x)}{\alpha(t,x)}\right)^2.
	\label{eq:app_full_GR_A_C}
\end{equation}
The scalar field contains odd harmonics, while the metric functions contain even harmonics. In the numerical implementation we truncate the Fourier series as
\begin{equation}
	\phi(t,x)
	=
	\phi_1(x)\cos(\omega t)
	+
	\phi_3(x)\cos(3\omega t),
	\label{eq:app_full_GR_phi_truncation}
\end{equation}
and
\begin{align}
	A(t,x)
	&=
	A_0(x)
	+
	A_2(x)\cos(2\omega t)
	+
	A_4(x)\cos(4\omega t),
	\label{eq:app_full_GR_A_truncation}
	\\
	C(t,x)
	&=
	C_0(x)
	+
	C_2(x)\cos(2\omega t)
	+
	C_4(x)\cos(4\omega t).
	\label{eq:app_full_GR_C_truncation}
\end{align}
The frequency \(\omega\) is an eigenvalue fixed by the boundary conditions. Selected configurations were also repeated with scalar harmonics
through \(\phi_5\) and metric harmonics through \(6\omega\) as a
convergence check. At \(|\tilde\lambda|=0\), the resulting maximum mass differs
from the production \(4\omega\) value by only about \(0.59\%\), and at
\(|\tilde\lambda|=4\) the difference is about \(0.2\%\). We therefore retain the
\(4\omega\) truncation for the production sequences and fitting analysis.

The time-dependent EKG equations are evaluated on a grid in the phase variable \(\omega t\), and the residuals are projected onto the retained cosine harmonics. This gives a radial boundary-value problem for the Fourier coefficients. Regularity at the origin requires
\begin{equation}
	\phi_j'(0)=0,
	\qquad
	A_0(0)=1,
	\qquad
	A_{j>0}(0)=0.
	\label{eq:app_full_GR_origin_bc}
\end{equation}
The central amplitude of the fundamental scalar harmonic,
\begin{equation}
	\phi_1(0),
	\label{eq:app_full_GR_central_amp}
\end{equation}
is used as the shooting or continuation parameter for constructing the solution sequence.

At the outer boundary, localization and asymptotic flatness require
\begin{equation}
	\phi_j(x_{\max})=0,
	\qquad
	C_0(x_{\max})=1,
	\qquad
	C_{j>0}(x_{\max})=0.
	\label{eq:app_full_GR_outer_bc}
\end{equation}
The functions \(A_j\) are determined by the radial equations and regularity conditions. In particular, the asymptotic behavior of \(A_0\) is used to extract the ADM mass.

The full-GR sequence is obtained by scanning over the central value \(\phi_1(0)\). The converged solution at one value is used as the initial guess for the next. This continuation method follows the nodeless branch of oscillaton-like solutions. A solution is rejected if the boundary-value solver fails to converge or if the fundamental harmonic \(\phi_1(x)\) develops nodes.

The mass is extracted from the zero mode \(A_0(x)\). In the convention used by the full-GR code, the cumulative mass is
\begin{equation}
	M_{\rm nr}(x)
	=
	\frac{x}{2}
	\left[
	1-\frac{1}{A_0(x)}
	\right].
	\label{eq:app_full_GR_mass_nonreduced}
\end{equation}
To compare with the reduced-Planck convention used in the EFT calculation, we use
\begin{equation}
	M(x)=8\pi M_{\rm nr}(x).
	\label{eq:app_full_GR_mass_reduced}
\end{equation}
As a consistency check, we also reconstruct the full time-dependent
fields from their retained Fourier harmonics and compute the
time-averaged mass profile directly from the scalar energy density.

For the maximum-mass configuration at
\(\tilde{\lambda}=-12\), the zero-mode metric prescription and the time-averaged density
prescription have corresponding fractional
differences approximately of order of \(10^{-6}\) in mass and
\(10^{-5}\) in radius. We therefore use the zero-mode metric
prescription for the full sequence, with the time-averaged density
calculation serving as a numerical consistency check.

The total soliton mass is read from the asymptotic value of \(M(x)\). The radius \(R_{95}\) is first found in the areal coordinate by imposing
\begin{equation}
	M(x<R_{95}^{\rm areal})=0.95M_s.
	\label{eq:app_full_GR_R95_areal}
\end{equation}
For comparison with the EFT solutions, which are written in isotropic coordinates, the areal radius is converted to an isotropic radius using the zero-mode radial metric. The transformation satisfies
\begin{equation}
	\frac{d\ln R_{\rm iso}}{dx}
	=
	\frac{\sqrt{A_0(x)}}{x}.
	\label{eq:app_full_GR_isotropic_transform}
\end{equation}
The integration constant is fixed by matching to the Schwarzschild relation at the outer boundary. The isotropic \(R_{95}\) is then obtained by applying the same enclosed-mass condition after the coordinate transformation.

The maximum mass in the full-GR sequence is identified from the scan over \(\phi_1(0)\). This provides the full relativistic reference curve against which the first- and second-order EFT mass--radius relations are compared.

\subsubsection{Harmonic convergence}
\label{app:harmonic_convergence}
\begin{figure}
    \centering
    \includegraphics[width=0.48\textwidth]{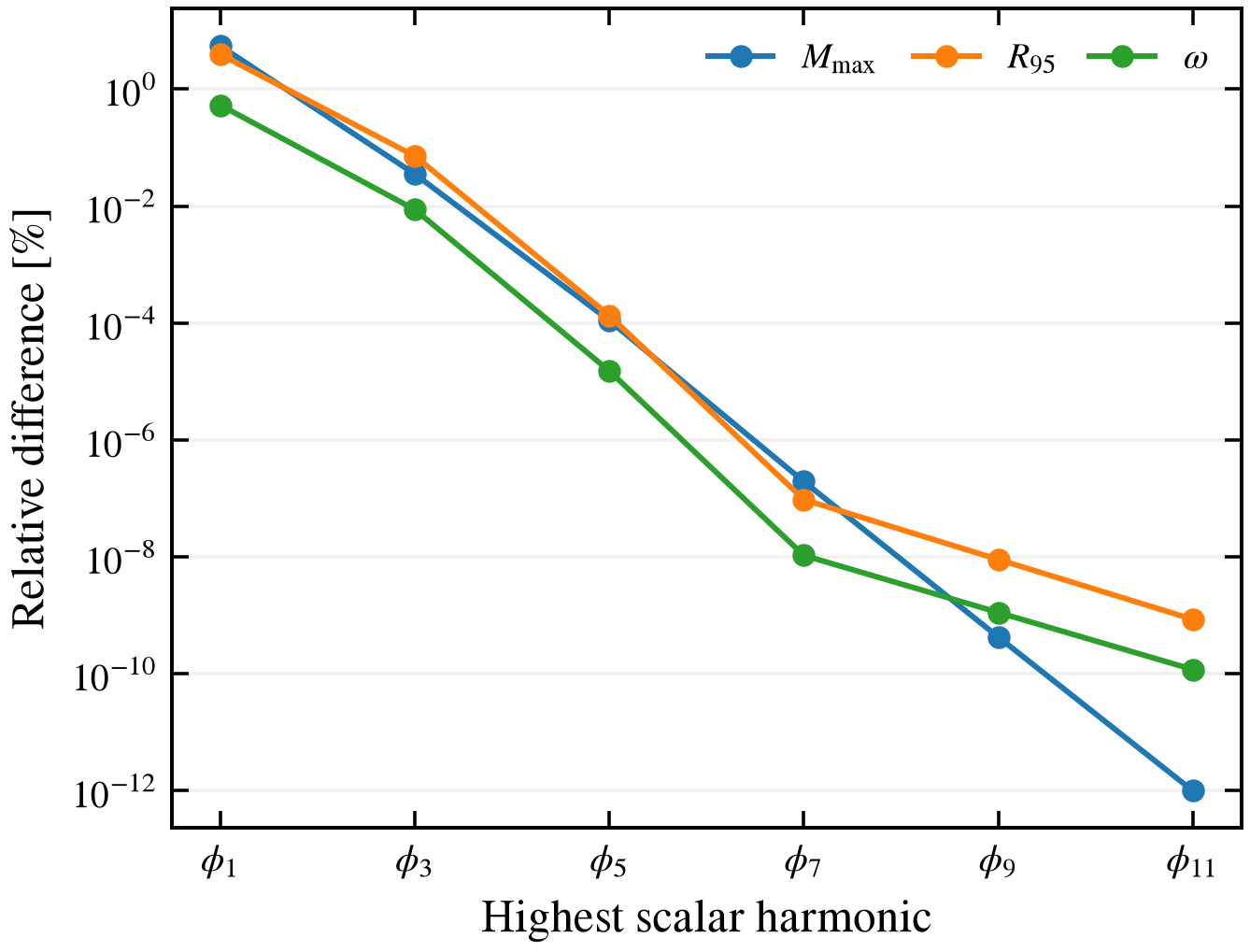}
    \caption{
Convergence of the full-cosine maximum-mass solution at
\(|\tilde{\lambda}|=12\) as the highest retained scalar harmonic is
increased. The metric Fourier basis is enlarged consistently through
the corresponding next even harmonic. The vertical axis shows the
relative difference from the highest-order result for
\(M_{\max}\), \(R_{95}\), and \(\omega\). The dominant correction is
obtained upon including \(\phi_3\); subsequent harmonics produce
rapidly decreasing changes.
}
    \label{fig:harmonic_observable_convergence}
\end{figure}

\begin{figure*}
    \centering
    \includegraphics[width=\textwidth]{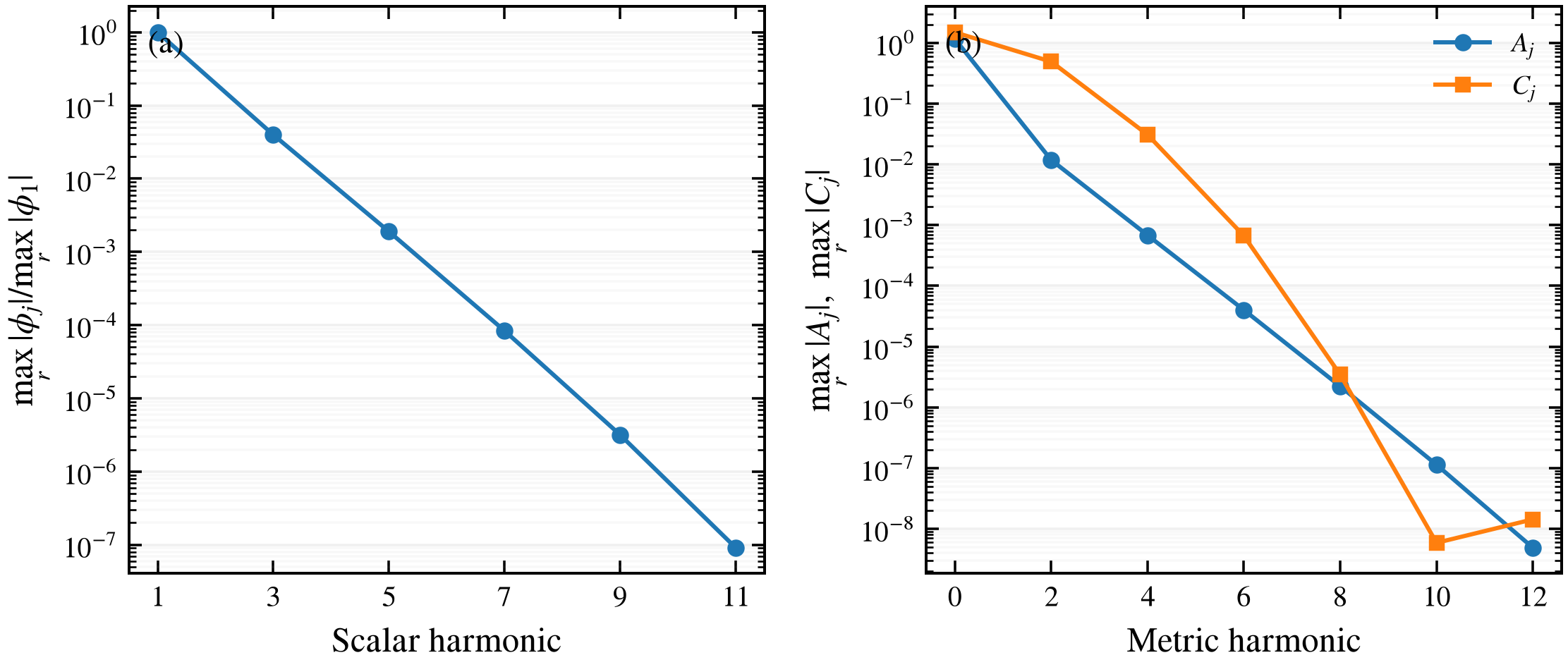}
    \caption{Fourier hierarchy of the converged full-cosine maximum-mass solution
at \(|\tilde{\lambda}|=12\). Panel (a) shows the scalar Fourier
coefficients normalized to the fundamental mode,
\(\max_r|\phi_j|/\max_r|\phi_1|\). Panel (b) shows the radial maxima
of the Fourier coefficients of the metric functions \(A\) and \(C\).
The higher modes are rapidly suppressed, demonstrating convergence of
the temporal Fourier expansion.}
    \label{fig:fourier_coefficient_convergence}
\end{figure*}

The full-GR results in the main text and in \ref{app:numerics_full_GR} use the \(4\omega\) production
truncation, containing the scalar modes \(\phi_1\) and \(\phi_3\) and
metric harmonics through \(4\omega\). Since higher harmonics can become
important when the real scalar explores the anharmonic region of the
axion potential \cite{Visinelli:2017ooc}, we test this truncation
explicitly.

For this purpose we use the complete cosine potential and successively
increase the Fourier basis as
\begin{equation}
\begin{split}
&\phi_1,\\
&\phi_1+\phi_3,\\
&\phi_1+\phi_3+\phi_5,\\
&\ldots,\\
&\phi_1+\phi_3+\phi_5+\phi_7+\phi_9+\phi_{11},
\end{split}
\end{equation}

while enlarging the metric basis consistently through
\(2\omega,4\omega,\ldots,12\omega\).
We perform this test at
\(|\tilde{\lambda}|=12\), where the maximum-mass configurations
reach one of the largest field excursions along the sequence and hence
provide a stringent test of the temporal truncation. 

Fig.~\ref{fig:harmonic_observable_convergence} shows the convergence
of \(M_{\max}\), \(R_{95}\), and the eigenfrequency \(\omega\), measured
relative to the highest-order calculation. The fundamental-only
solution exhibits a visible truncation error. Including the third
scalar harmonic removes nearly all of this difference, while the
successive corrections from \(\phi_5\), \(\phi_7\), and higher modes
decrease rapidly. Thus, although a single-frequency approximation is
not sufficiently accurate for this configuration, the \(4\omega\)
production truncation already captures the dominant higher-harmonic
correction.

The convergence of the integrated observables is accompanied by a
rapid suppression of the Fourier coefficients themselves.
Fig.~\ref{fig:fourier_coefficient_convergence} shows the harmonic
content of the highest-order solution. The Fourier coefficients of the metric functions \(A\) and \(C\)
display a similarly strong overall suppression with increasing
harmonic order. We therefore used the \(4\omega\) truncation for the production
quartic sequences in the main analysis, while the higher-order
calculations are used as an explicit convergence test. {This convergence establishes that the retained Fourier basis accurately
describes the periodic core solution and its structural observables. It
does not determine the amplitude of a possible outgoing radiative tail
or the lifetime of the corresponding real-field configuration, which
would require a time-dependent calculation with radiative boundary
conditions.}

\subsubsection{Mass and Radius Fits}
\label{app:fit}

We denote the absolute value of the dimensionless attractive self-interaction parameter by $|\tilde{\lambda}|$.
All fits in this appendix are calibrated over
\(4\leq|\tilde{\lambda}|\leq700\). For reference, representative raw numerical maximum-mass points
used in the comparison are listed in
Table~\ref{tab:raw_max_points}. For the full-GR solutions, $R_{95}$ is reported
in isotropic coordinates for direct comparison with the SP and EFT
results.
\begin{table*}[t]
\caption{
Representative raw numerical maximum-mass configurations used in the
comparison of the four descriptions. The values are taken directly
from the numerical turning points and are not obtained from the
interpolation fits. Masses and radii are given in the dimensionless
conventions used throughout the text. For the full-GR solutions,
$R_{95}$ is reported in isotropic coordinates for direct comparison
with the SP and EFT results.
}
\label{tab:raw_max_points}
\begin{ruledtabular}
\begin{tabular}{c cc cc cc cc}
& \multicolumn{2}{c}{SP}
& \multicolumn{2}{c}{1st-order EFT}
& \multicolumn{2}{c}{2nd-order EFT}
& \multicolumn{2}{c}{Full GR} \\
$|\tilde{\lambda}|$
& $M_{\max}$ & $R_{95,\max}$
& $M_{\max}$ & $R_{95,\max}$
& $M_{\max}$ & $R_{95,\max}$
& $M_{\max}$ & $R_{95,\max}$ \\
\hline
4   & 25.4464 & 4.1792  & 16.0979 & 3.3273
    & 14.1666 & 5.0842  & 13.6727 & 5.9468 \\
10  & 16.0942 & 6.6936  & 12.4211 & 6.4401
    & 11.8725 & 6.9730  & 11.7452 & 7.2299 \\
20  & 11.3804 & 9.4282  &  9.7746 & 9.3193
    &  9.6108 & 9.6743  &  9.5888 & 9.5224 \\
40  &  8.0472 & 13.2981 &  7.4004 & 13.3397
    &  7.3366 & 13.0295 &  7.3566 & 13.0429 \\
100 &  5.0896 & 21.1597 &  4.9106 & 20.9817
    &  4.9051 & 20.9879 &  4.9054 & 20.9856 \\
700 &  1.9236 & 55.7517 &  1.9132 & 56.1128
    &  1.9133 & 55.4571 &  1.9133 & 55.5573 \\
\end{tabular}
\end{ruledtabular}
\end{table*}
For the Newtonian Schrödinger--Poisson sequence, the maximum mass follows the dilute attractive self-interaction scaling. We therefore fit the NR sequence with a power law,
\begin{equation}
    M_{\max}^{\rm NR}(|\tilde{\lambda}|)
    =
    A_{\rm NR}|\tilde{\lambda}|^{-1/2},
    \label{eq:Mmax_NR_powerlaw}
\end{equation}
with
\begin{equation}
    A_{\rm NR}=50.893 .
\end{equation}

For the relativistic correction and full GR sequences, we fit the maximum mass using the interpolation form
\begin{equation}
    M_{\max}^{X}(|\tilde{\lambda}|)
    =
    \frac{M_0^X}
    {\sqrt{1+C_X|\tilde{\lambda}|}},
    \qquad
    C_X=
    \left(\frac{M_0^X}{A_{\rm NR}}\right)^2 .
    \label{eq:Mmax_fit_lambda}
\end{equation}
This form approaches a finite value \(M_0\) in the weak coupling limit and recovers the expected large-\(|\tilde{\lambda}|\) behavior
\begin{equation}
    M_{\max}^{X}(|\tilde{\lambda}|)
    \simeq
    A_{\rm NR}|\tilde{\lambda}|^{-1/2},
    \qquad |\tilde{\lambda}|\to\infty .
\end{equation}
For the radius, we use a pivoted square root scaling,
\begin{equation}
    R_{95,\max}^{X}(|\tilde{\lambda}|)
    =
    R_\star^X
    \sqrt{
        1+
        \left(
            \frac{B_{\rm NR}}{R_\star^X}
        \right)^2
        (|\tilde{\lambda}|-|\tilde{\lambda}|_\star)
    }
    \label{eq:Rmax_fit_lambda_pivot}
\end{equation}
with 
\begin{equation}
     B_{\rm NR}=2.106,\qquad |\tilde{\lambda}|_\star=100.
\end{equation}

Thus \(D_X=(B_{\rm NR}/R_\star^X)^2\), and every radius fit has the
common asymptotic limit
\(R_{95,\max}^{X}\simeq B_{\rm NR}|\tilde{\lambda}|^{1/2}\).

The fitted maximum mass relations are
\begin{align}
    M_{\max}^{\rm NR}(|\tilde{\lambda}|)
    &=
    50.893\,|\tilde{\lambda}|^{-1/2},
    \nonumber\\
    M_{\max}^{\rm 1st\,Order\,EFT}(|\tilde{\lambda}|)
    &=
    \frac{20.037}
    {\sqrt{1+0.155\,|\tilde{\lambda}|}},
    \nonumber\\
    M_{\max}^{\rm 2nd\,Order\,EFT}(|\tilde{\lambda}|)
    &=
    \frac{17.395}
    {\sqrt{1+0.117\,|\tilde{\lambda}|}},
    \nonumber\\
    M_{\max}^{\rm GR,\,Real}(|\tilde{\lambda}|)
    &=
    \frac{16.907}
    {\sqrt{1+0.110\,|\tilde{\lambda}|}}.
    \label{eq:fitted_mass_relations_updated}
\end{align}

The corresponding pivoted radius fits, \(|\tilde{\lambda}|_\star=100\), are
\begin{align}
    R_{95,\max}^{\rm NR}(|\tilde{\lambda}|)
    &=
    21.060
    \sqrt{1+0.01000(|\tilde{\lambda}|-100)},
    \nonumber\\
    R_{95,\max}^{\rm 1st\,Order\,EFT}(|\tilde{\lambda}|)
    &=
    20.979
    \sqrt{1+0.01008(|\tilde{\lambda}|-100)},
    \nonumber\\
    R_{95,\max}^{\rm 2nd\,Order\,EFT}(|\tilde{\lambda}|)
    &=
    21.132
    \sqrt{1+0.00993(|\tilde{\lambda}|-100)},
    \nonumber\\
    R_{95,\max}^{\rm GR,\,Real}(|\tilde{\lambda}|)
    &=
    21.373
    \sqrt{1+0.00971(|\tilde{\lambda}|-100)}.
    \label{eq:fitted_radius_relations_updated}
\end{align}

Eliminating \(|\tilde{\lambda}|\) gives the corresponding mass--radius relation for the maximum mass sequence. 
\begin{equation}
    M_{\max}^{\rm fit}(R)
    =
    \frac{M_0}
    {
    \sqrt{
    1+C|\tilde{\lambda}|_\star
    +
    \frac{C}{D}
    \left[
    \left(\frac{R}{R_\star}\right)^2-1
    \right]
    }
    }.
    \label{eq:mass_radius_fit_pivot_compact}
\end{equation}

In the large-\(|\tilde{\lambda}|\) limit, the fitted relations imply
\begin{equation}
    M_{\max}R_{95,\max}\rightarrow {\rm constant}.
\end{equation}
The asymptotic products obtained from the updated fits are approximately
\begin{equation}
    M_{\max}^{X}(|\tilde{\lambda}|)
    R_{95,\max}^{X}(|\tilde{\lambda}|)
    \xrightarrow[|\tilde{\lambda}|\to\infty]{}
    A_{\rm NR}B_{\rm NR}
    \simeq107.18 ,
\end{equation}
We finally emphasize that the fitting expressions in this appendix are
empirical interpolations calibrated over \(4\leq|\tilde{\lambda}|\leq700\).
The fitted parameters \(M_0^X\) are therefore not estimates of the
physical \(|\tilde{\lambda}|=0\) maximum masses. In particular, the full-GR
zero-coupling limit is determined directly from the oscillaton sequence,
for which the refined \(6\omega\) calculation gives
\(M_{\max}^{\rm GR}(0)/(8\pi)=0.606\,M_P^2/m\).

\bibliography{axionstar}
\end{document}